\documentclass[aps,pra,amsmath,twocolumn,amssymb,superscriptaddress]{revtex4-1}

\usepackage{amssymb,amsfonts,amsmath}
\usepackage{color}
\usepackage[apple mac]{inputenc}
\usepackage[english]{babel}
\usepackage{times}
\usepackage{latexsym}
\usepackage{fancyhdr}
\usepackage{verbatim}
\usepackage{graphicx}
\usepackage{epsf}
\usepackage{subfigure}
\usepackage{bm}
\usepackage{epstopdf}

\definecolor{Nathanblue}{rgb}{0.,0.24,0.51}
\newcommand{\blue}{\color{Nathanblue}}

\definecolor{orange}{rgb}{0.96,0.24,0.00}

\usepackage{mathtools}

\def\be{\begin{equation}}
\def\ee{\end{equation}}

\begin{document}

\title{{\blue Floquet engineering of optical nonlinearities:~a quantum many-body approach}}

\author{N. Goldman}
\email[]{ngoldman@ulb.ac.be}
\affiliation{CENOLI,
Universit\'e Libre de Bruxelles, CP 231, Campus Plaine, B-1050 Brussels, Belgium}

\begin{abstract}
Subjecting a physical system to a time-periodic drive can substantially modify its properties and applications. This Floquet-engineering approach has been extensively applied to a wide range of classical and quantum settings in view of designing synthetic systems with exotic properties. Considering a general class of two-mode nonlinear optical devices, we show that effective optical nonlinearities can be created by subjecting the light field to a repeated pulse sequence, which couples the two modes in a fast and time-periodic manner. The strength of these drive-induced optical nonlinearities, which include an emerging four-wave mixing, can be varied by simply adjusting the pulse sequence. This leads to topological changes in the system's phase space, which can be detected through light intensity and phase measurements. Our proposal builds on an effective-Hamiltonian approach, which derives from a parent quantum many-body Hamiltonian describing driven interacting bosons. As a corollary, our results equally apply to Bose-Einstein condensates in driven double-well potentials, where pair tunneling effectively arises from the periodic pulse sequence. Our scheme offers a practical route to engineer and finely tune exotic nonlinearities and interactions in photonics and ultracold quantum gases.
\end{abstract}

\date{\today}

\maketitle

\section{Introduction}

Floquet engineering is a vaste and pluridisciplinary program, which consists in controlling and designing synthetic systems with time-periodic drives in view of exploring novel phenomena~\cite{eckardt2017colloquium,oka2019floquet,rudner2020band,weitenberg2021tailoring}. This general approach concerns a wide range of physical platforms, including ultracold quantum gases~\cite{eckardt2017colloquium,weitenberg2021tailoring}, solid-state materials~\cite{oka2019floquet,rudner2020band}, universal quantum simulators and computers~\cite{georgescu2014quantum,altman2021quantum}, mechanical~\cite{salerno2016floquet} and acoustical~\cite{fleury2016floquet} systems, and photonic devices~\cite{rechtsman2013photonic,schine2016synthetic,roushan2017chiral,ozawa2019topological}. 

More specifically, Floquet engineering can be applied to modify the band structure of lattice systems~\cite{eckardt2017colloquium,rudner2020band}, generate artificial gauge fields~\cite{goldman2014periodically,aidelsburger2018artificial} and design complex interaction processes~\cite{rapp2012ultracold,ajoy2013quantum,di2014quantum,daley2014effective,hung2016quantum,pieplow2018generation,lee2018floquet,choi2020robust,barbiero2020bose,dehghani2021light,geier2021floquet}. These remarkable possibilities open the door to the experimental exploration of a broad range of intriguing physical phenomena, such as high-temperature superconductivity~\cite{fausti2011light,mitrano2016possible}, magnetism~\cite{struck2011quantum,struck2013engineering,gorg2018enhancement}, topological physics~\cite{ozawa2019topological,rudner2020band,weitenberg2021tailoring}, many-body localization~\cite{ponte2015many,abanin2019colloquium}, chaos-assisted tunneling~\cite{hensinger2001dynamical,arnal2020chaos}, and lattice gauge theories~\cite{barbiero2019coupling,schweizer2019floquet}.

Floquet engineering has been particularly fruitful in the realm of photonics, where various settings and periodic-driving scenarios have been proposed and experimentally realized. In laser-written optical waveguide arrays~\cite{szameit2010discrete}, where waveguides can be finely modulated along the propagation direction, Floquet schemes were implemented in view of generating topological band structures~\cite{rechtsman2013photonic,mukherjee2017experimental,maczewsky2017observation,mukherjee2018state,mukherjee2020observation,mukherjee2021observation}, synthetic dimensions~\cite{lustig2019photonic} and artificial magnetic fields~\cite{mukherjee2018experimental} for light; these settings led to the observation of dynamic localization~\cite{longhi2006observation,szameit2009inhibition}, coherent destruction of tunneling~\cite{della2007visualization} and modulation-assisted tunneling~\cite{mukherjee2015modulation}, Floquet anomalous edge states~\cite{mukherjee2017experimental,maczewsky2017observation}, disorder-induced topological states~\cite{stutzer2018photonic},  topological solitons~\cite{mukherjee2020observation,mukherjee2021observation}, as well as Aharonov-Bohm cages~\cite{mukherjee2018experimental}. In the context of optical resonators, electro-optical modulators were used to resonantly couple different cavity modes and realize synthetic dimensions~\cite{yuan2018synthetic,dutt2019experimental,dutt2020single,balvcytis2021synthetic,englebert2021bloch}, while non-planar geometries were designed to create stroboscopic dynamics reflecting an effective magnetic field for photons~\cite{schine2016synthetic}. In the context of circuit-QED, time-modulated couplers connecting superconducting qubits were exploited to create artificial magnetic fields for strongly-interacting photons hopping on a lattice~\cite{roushan2017chiral}. Finally, drive-induced optical nonlinearities recently emerged as an exciting avenue in the context of polaritons~\cite{clark2019interacting,johansen2020multimode},  insulating materials~\cite{shan2021giant}, and high-Q microwave cavities coupled to transmon qubits~\cite{zhang2022drive}.

In this work, we explore the possibility of generating effective optical nonlinearities upon subjecting a propagating light field to a repeated pulse sequence. The emergence of effective nonlinearities is investigated for a general class of two-mode nonlinear systems, described by the nonlinear Schr\"odinger equation, where the driving sequence corresponds to a succession of fast (linear) mode-mixing operations. This framework captures a broad range of nonlinear physical settings, including two-mode optical cavities~\cite{cao2017experimental,hill2020effects,garbin2020asymmetric}, optical waveguide couplers~\cite{szameit2010discrete,szameit2009inhibition} and coupled superconducting microwave cavities~\cite{roushan2017chiral}, but also quantum gases trapped in double-well potentials~\cite{andrews1997observation,smerzi1997quantum} and two-component Bose-Einstein condensates~\cite{zibold2010classical}. In fact, the periodically-driven nonlinear Schr\"odinger equation (or Gross-Pitaevskii equation~\cite{pitaevskii2016bose}) was previously investigated in the context of driven quantum gases~\cite{holthaus2001towards,kramer2005parametric,susanto2008effects,lellouch2017parametric,driben2018nonlinearity,kidd2019quantum}, and more recently, in the realm of topological photonics~\cite{mukherjee2020observation,maczewsky2020nonlinearity,mukherjee2021observation,mochizuki2021fate,ivanov2021topological}; see also Refs.~\cite{szameit2009inhibition,higashikawa2018floquet,sentef2020quantum}. 

\begin{figure}[h!]
\includegraphics[width = \linewidth]{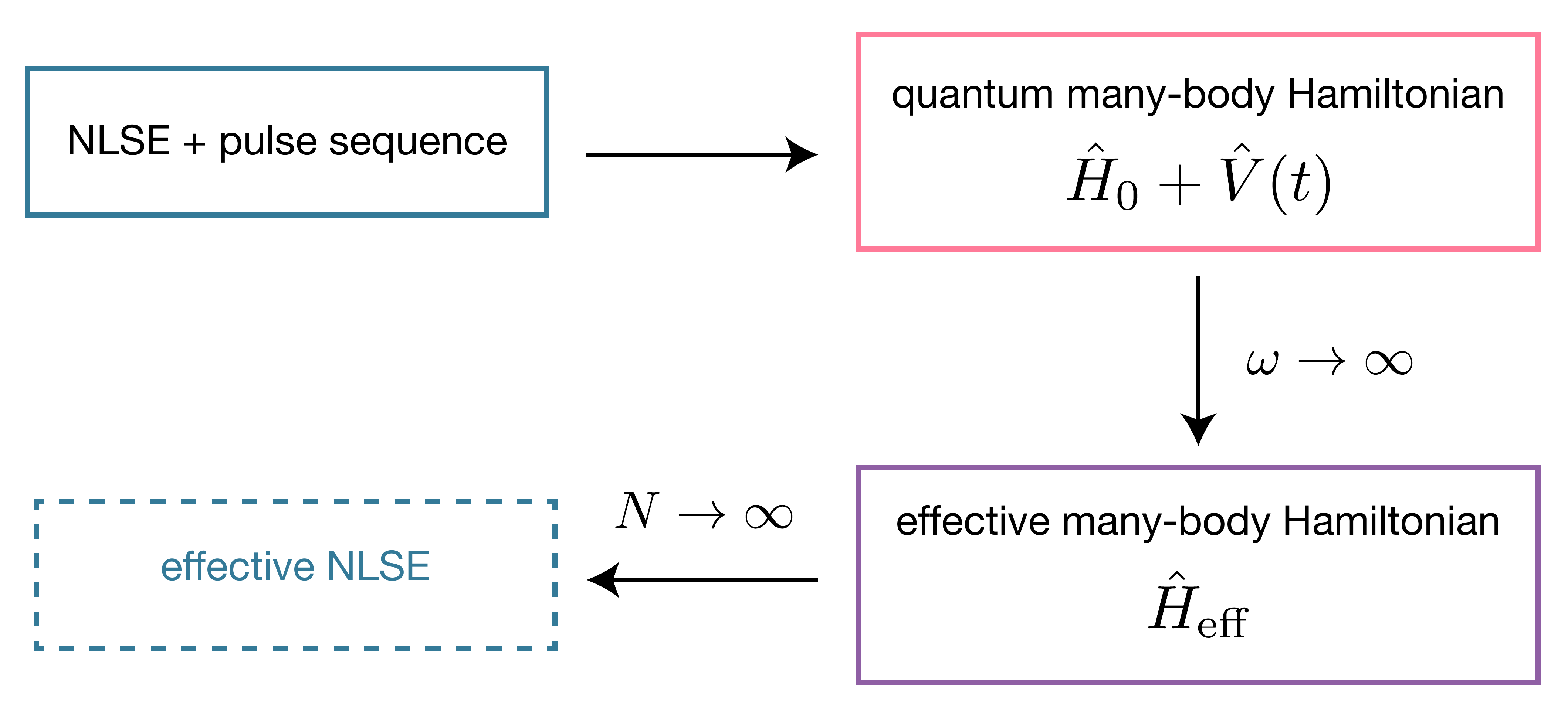}
\caption{Schematic of the approach. We consider a general class of two-mode nonlinear systems, driven by a periodic pulse sequence and described by the nonlinear Schr\"odinger equation (NLSE). To analyse these settings, we introduce a parent quantum many-body Hamiltonian, $\hat H_0 + \hat V(t)$, which describes two species of interacting bosons driven by a periodic pulse sequence. From this, we derive an effective quantum Hamiltonian $\hat H_{\text{eff}}$ in the high-frequency limit of the drive ($\omega\!\rightarrow\infty$). We then derive the effective nonlinear Schr\"odinger equation upon taking the classical limit, $N\!\rightarrow\infty$, where $N$ is the number of bosons, hence revealing the effective nonlinearities generated by the driving pulse sequence.
}
\label{fig_approach}
\end{figure}

Our analysis is based on a parent quantum many-body Hamiltonian, which describes two species of interacting bosons subjected to a periodic pulse sequence; see the sketch in Fig.~\ref{fig_approach}. This theoretical framework captures the physics of driven nonlinear optical settings in the classical (mean-field) limit~\cite{pitaevskii2016bose,carusotto2013quantum,cao2020reconfigurable}. Specifically, we first derive an effective quantum Hamiltonian that well describes the stroboscopic dynamics of the driven parent quantum system, in the high-frequency regime of the pulse sequence~\cite{goldman2014periodically,goldman2015periodically,bukov2015universal,eckardt2015high,mikami2016brillouin}. From this, we then derive the effective classical equations of motion, hence revealing the effective optical nonlinearities generated by the driving sequence [Section~\ref{section_quantum}]. We explore the validity of both approximations (i.e.~the high-frequency approximation related to the drive and the mean-field approximation associated with the classical limit) through numerical simulations of the quantum and classical dynamics, comparing the full time dynamics generated by the pulse sequence to the effective descriptions [Section~\ref{section_numerics}]. As a by-product, this analysis illustrates how the effective nonlinearities can be detected through the dynamics of simple observables:~the relative intensity and relative phase of the two optical modes. We then discuss how the strength of the effective optical nonlinearities can be tuned by simply adjusting the pulse sequence [Section~\ref{sect_imbalanced}]. This control over drive-induced optical nonlinearities is directly reflected in the topology of the system’s phase space, which can be detected through light intensity and phase measurements.

\section{The driven nonlinear system and its effective description}

We consider a class of two-mode nonlinear systems, described by the (possibly coupled) nonlinear Schr\"odinger equations
\begin{align}
&i \frac{\partial \psi_1}{\partial t} = \left ( - \gamma \frac{\partial^2}{\partial x^2} + \vert \psi_1 \vert^2 + \beta \vert \psi_2 \vert^2   \right ) \psi_1 - \frac{\Omega_0}{2} \psi_2, \notag \\
&i \frac{\partial \psi_2}{\partial t} = \left ( -\gamma \frac{\partial^2}{\partial x^2} + \vert \psi_2 \vert^2 + \beta \vert \psi_1 \vert^2   \right ) \psi_2 - \frac{\Omega_0}{2} \psi_1. \label{NLS}
\end{align}
Here, $\psi_{1,2}(x,t)$ denote the complex amplitude of the fields corresponding to the modes $\sigma=1,2$; they depend on the evolution ``time" $t$ and the ``spatial" coordinate $x$. The focus of this work is set on the ``internal" dynamics associated with the two modes, such that the ``spatial" coordinate $x$ [and the related kinetic-energy term~$\sim \gamma$ in Eq.~\eqref{NLS}] does not play any role in the following. For the sake of generality, the equations of motion~\eqref{NLS} contain two types of nonlinearities, which are generically present in optical cavities~\cite{cao2017experimental,hill2020effects,garbin2020asymmetric}:~the so-called self-phase modulation and the cross-phase modulation, whose respective strengths are set by the parameter $\beta$; we have also included a static linear coupling of strength $\Omega_0/2$. We point out that the nonlinear equations~\eqref{NLS} are decoupled in the limit $\Omega_0\!=\!\beta\!=\!0$, i.e.~in the absence of linear coupling and cross-phase modulation. 

While Eq.~\eqref{NLS} naturally describes the two polarization modes of a light field propagating in a lossless cavity~\cite{cao2017experimental,hill2020effects,garbin2020asymmetric}, or light propagating in a pair of adjacent waveguides~\cite{szameit2010discrete,szameit2009inhibition}, it should be noted that Eq.~\eqref{NLS}  also captures the physics of quantum gases trapped in a double well potential and two-component Bose-Einstein condensates~\cite{smerzi1997quantum,zibold2010classical}. 

In order to create effective nonlinearities in Eq.~\eqref{NLS}, we now include a time-periodic pulse sequence of period $T$, which mixes the two modes in a stroboscopic manner:~
\begin{itemize}
\item Pulse $\oplus$: At times $t_{\frak{n}}^{\oplus}\!=\!(T/2)\!\times\!(2\frak{n}-1)$, where $\frak{n}\!\in\!\mathbb{N}$, the two components undergo the mixing operation
\begin{align}\label{eq_exchange}
\psi_1 \rightarrow (1/\sqrt{2}) \left (\psi_1 +i \psi_2 \right ), \,\, \psi_2 \rightarrow (1/\sqrt{2}) \left (i \psi_1 + \psi_2 \right).
\end{align}
\item Pulse $\ominus$: at times $t_{\frak{n}}^{\ominus}\!=\! T\!\times\!\frak{n}$,  the system undergoes the reverse  operation
\begin{align}\label{eq_exchange_2}
\psi_1 \rightarrow (1/\sqrt{2}) \left (\psi_1 -i \psi_2 \right ), \,\, \psi_2 \rightarrow (1/\sqrt{2}) \left (\psi_2 -i \psi_1\right).
\end{align}
\end{itemize}

\begin{figure}[h!]
\includegraphics[width = \linewidth]{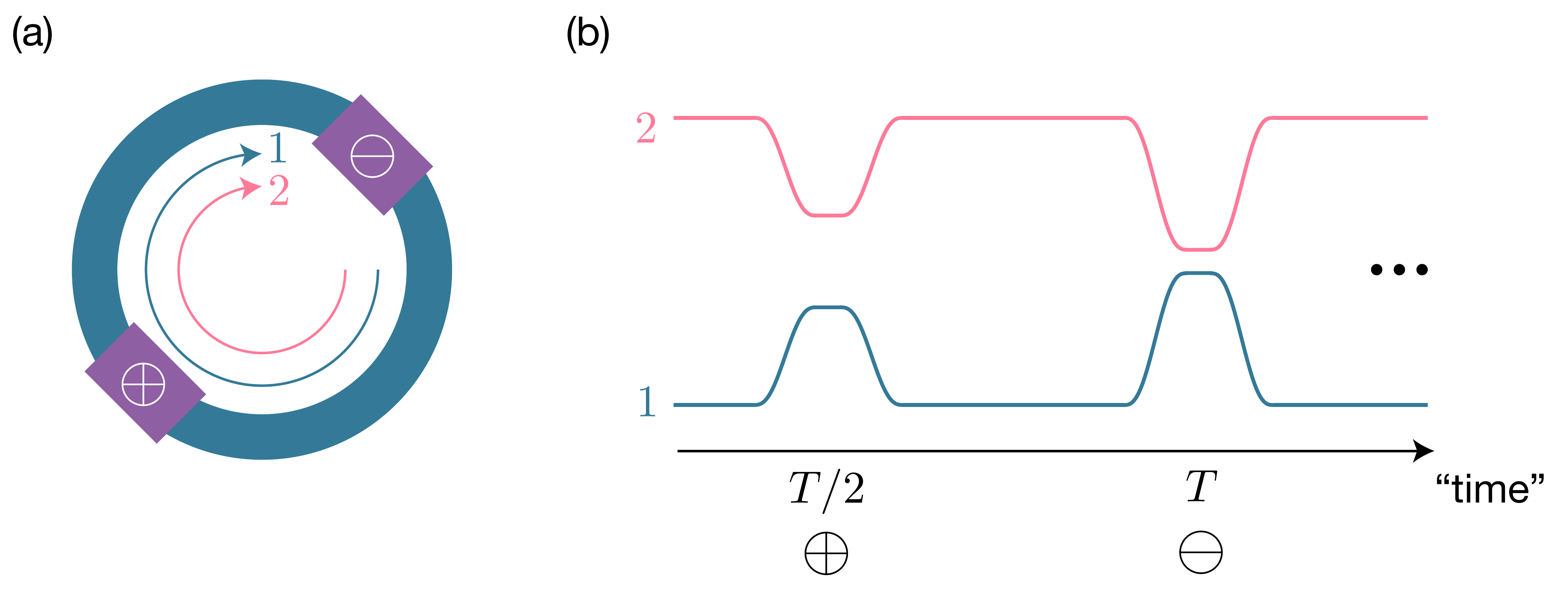}
\caption{Two possible realizations in optics: (a) Two modes in an optical ring cavity ($1$ and $2$), repeatedly undergoing  mixing operations ($\oplus$ and $\ominus$) along the ring. These operations correspond to a coupling between the two polarization eigenmodes of the cavity, as realized by means of quarter-wave plates; see Eqs.~\eqref{eq_exchange}-\eqref{eq_exchange_2}. (b) Two optical waveguides ($1$ and $2$) with modulated inter-waveguide separation, realizing a ``time-periodic" linear coupling between the two optical modes [Eq.~\eqref{NLS_time}]. In both cases, the ``time" coordinate corresponds to the propagation direction~\cite{szameit2010discrete,fatome2021self}.
}
\label{fig_sketch}
\end{figure}

In a two-mode optical cavity~\cite{cao2017experimental,hill2020effects,garbin2020asymmetric}, these pulsed operations would correspond to a coupling between the two polarization eigenmodes of the cavity, as directly realized by means of quarter-wave plates~\cite{kockaert2006fast,kozyreff2006fast}; see the sketch in Fig.~\ref{fig_sketch}(a). In this case, the ``time" coordinate $t$ should be interpreted as the propagation distance along the cavity~\cite{fatome2021self}. 

\begin{figure}[h!]
\includegraphics[width = \linewidth]{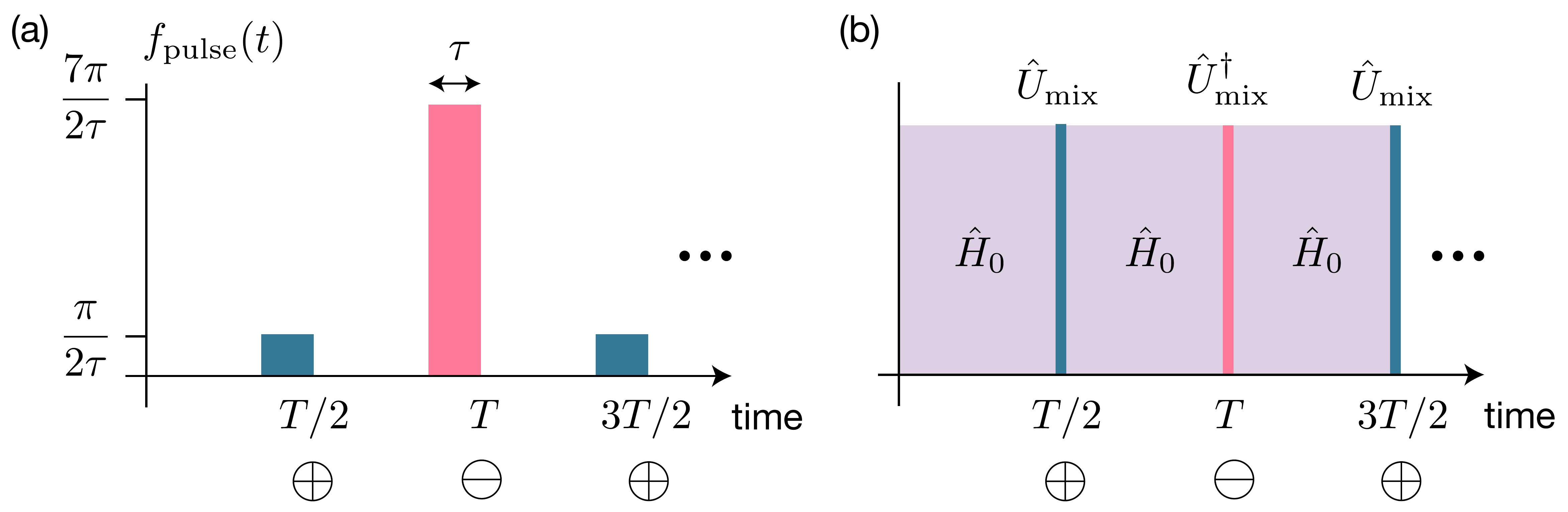}
\caption{(a) Pulse function determining the stroboscopic activation of the linear coupling in Eqs.~\eqref{NLS_time}-\eqref{pulse}. (b) The pulse sequence associated with the time-evolution operator in Eq.~\eqref{U_sequence}, which involves stroboscopic mixing operations $\hat U_{\text{mix}}^{(\dagger)}$ separated by ``free" time evolution ($\hat H_0$).
}
\label{fig1}
\end{figure}

More generally, when the mixing processes in Eqs.~\eqref{eq_exchange}-\eqref{eq_exchange_2} cannot be directly performed by a device, they can also be realized by activating a linear coupling between the two modes, during a short pulse duration $\tau \ll T$, such that the equations of motion of the driven system can be written in the form
\begin{align}
&i \frac{\partial \psi_1}{\partial t} = \left ( - \gamma \frac{\partial^2}{\partial x^2} + \vert \psi_1 \vert^2 + \beta \vert \psi_2 \vert^2   \right ) \psi_1 - \frac{\Omega (t)}{2} \psi_2, \label{NLS_time}\\
&i \frac{\partial \psi_2}{\partial t} = \left ( -\gamma \frac{\partial^2}{\partial x^2} + \vert \psi_2 \vert^2 + \beta \vert \psi_1 \vert^2   \right ) \psi_2 - \frac{\Omega (t)}{2} \psi_1. \notag
\end{align}
Here, the function $\Omega(t)\!=\!\Omega_0 - f_{\text{pulse}}(t)$ includes the pulse sequence defined by the function [Fig.~\ref{fig1}(a)]
\begin{align}
f_{\text{pulse}}(t) &= (\pi/2\tau) \qquad  t_{\frak{n}}^{\oplus}-\tau \le t \le t_{\frak{n}}^{\oplus}, \notag\\
&= (7\pi/2\tau) \hspace{0.54cm}  t_{\frak{n}}^{\ominus}-\tau \le t \le t_{\frak{n}}^{\ominus}, \notag\\
&=0 \hspace{1.55cm} \text{otherwise} . \label{pulse}
\end{align}
To verify that the drive in Eqs.~\eqref{NLS_time}-\eqref{pulse} indeed realizes the mixing operations in Eqs.~\eqref{eq_exchange}-\eqref{eq_exchange_2}, we restrict ourselves to the (linear) driving terms in the coupled Schrödinger equations~\eqref{NLS_time} and we obtain the time-evolution operators corresponding to the first and second pulses, respectively:
\begin{align}
&\hat U (t_{\frak{n}}^{\oplus}; t_{\frak{n}}^{\oplus} - \tau)=e^{-i \tau (\frac{\pi}{2 \tau})(-\frac{1}{2}) \hat \sigma_x}=e^{i \frac{\pi}{4} \hat \sigma_x}\equiv \hat U_{\text{mix}},\label{pi_over_two}\\
&\hat U (t_{\frak{n}}^{\ominus}; t_{\frak{n}}^{\ominus} - \tau)=e^{-i \tau (\frac{7\pi}{2 \tau})(-\frac{1}{2}) \hat \sigma_x}=e^{-i \frac{\pi}{4} \hat \sigma_x} = \hat U_{\text{mix}}^{\dagger},\notag
\end{align}
where $\hat \sigma_x$ is the standard Pauli matrix. The operators $\hat U_{\text{mix}}$ and $\hat U_{\text{mix}}^{\dagger}$ in Eq.~\eqref{pi_over_two} indeed realize the mixing operations in Eqs.~\eqref{eq_exchange}-\eqref{eq_exchange_2}, respectively. We note that these mixing operations are reminiscent of $\pi/2$ pulses in quantum optics~\cite{gardiner2014quantum,kitagawa1993squeezed,gross2010nonlinear}. We also point out that the choice of the pulse function in Eq.~\eqref{pulse} is not unique, e.g.~the amplitude of the second pulse ($\ominus$) can be set to the value $(-\pi/2 + 2 \pi \mathfrak{p})/\tau$, with $\mathfrak{p}\!\in\!\mathbb{Z}$, without affecting the dynamics. The advantage of the pulse function proposed in Eq.~\eqref{pulse} is that the linear coupling does not change sign over time, which can be convenient for certain physical realizations.

In optical-waveguides settings~\cite{szameit2010discrete}, the two modes ($1$ and $2$) would describe light propagating in two adjacent waveguides. In this case, the pulsed linear couplings in Eqs.~\eqref{NLS_time}-\eqref{pulse} can be realized by abruptly changing the spatial separation between the two waveguides; see Fig.~\ref{fig_sketch}(b) for a sketch and Refs.~\cite{mukherjee2017experimental,maczewsky2017observation,mukherjee2020observation,mukherjee2021observation} for experimental realizations using ultrafast-laser-inscribed waveguides. Such optical-waveguides settings could benefit from the state-recycling technique of Refs.~\cite{mukherjee2018state,duncan2020synthetic}, where light is re-injected into the waveguides (and possibly modified) at every roundtrip; see also Refs.~\cite{kraych2019nonlinear,kraych2019statistical,kraych2020instabilites} regarding setups based on recirculating fiber loops. 

In the context of quantum gases trapped in a double well potential, the linear coupling between the two neighboring sites (orbitals) can be activated in a pulsed manner through a dynamical variation of the potential barrier~\cite{schumm2005bose}; see also Ref.~\cite{gross2010nonlinear}, where fast $\pi/2$ pulses were implemented in a two-component Bose-Einstein condensate.

In the limit of a fast pulse sequence, namely, when the period of the drive $T\ll T_{\text{eff}}$ is much smaller than the effective ``time" scale of the system (to be discussed below), the stroboscopic time-evolution of the nonlinear system is found to be well described by the effective equations of motion
\begin{align}
&i \frac{\partial \psi_1}{\partial t} = \left ( - \gamma \frac{\partial^2}{\partial x^2} + \frac{\chi}{4} \vert \psi_1 \vert^2    \right ) \psi_1 - \frac{\Omega_0}{2} \psi_2 - \frac{\chi}{4} \psi_1^* \psi_2^2, \notag\\
&i \frac{\partial \psi_2}{\partial t} = \left ( -\gamma \frac{\partial^2}{\partial x^2} + \frac{\chi}{4} \vert \psi_2 \vert^2    \right ) \psi_2 - \frac{\Omega_0}{2} \psi_1 - \frac{\chi}{4} \psi_2^* \psi_1^2 , \label{NLS_effective}
\end{align}
where we introduced the quantity $\chi\!=\!1-\beta$, and where the system is assumed to be measured stroboscopically ($t\!=\! T\!\times\!\frak{n}$, with $\frak{n}\!\in\!\mathbb{N}$). Comparing Eq.~\eqref{NLS_effective} with the original Eq.~\eqref{NLS}, we find that the repeated mixing processes in Eqs.~\eqref{eq_exchange}-\eqref{eq_exchange_2} effectively produce a new form of nonlinearity, commonly known in optics as four-wave mixing~\cite{agrawal2001applications,agrawal2012nonlinear}. The drive also renormalizes the self-phase-modulation and it effectively annihilates the cross-phase modulation. We point out that the effective four-wave mixing is induced even in the limit of two initially decoupled modes ($\beta\!=\!\Omega_0\!=\!0$).

It is the aim of the following Sections~\ref{section_quantum}-\ref{section_numerics} to demonstrate the effective description displayed in Eq.~\eqref{NLS_effective} and to explore its regimes of validity. We then introduce an ``imbalanced" pulse sequence in Section~\ref{sect_imbalanced}, which allows one to tune the relative strengths of effective nonlinearities and induce topological changes in phase space. While this work sets the focus on the Floquet engineering of classical nonlinear systems, the results obtained below also apply to the quantum dynamics of driven interacting bosonic systems.

As a technical note, we point out that the mixing processes in Eqs.~\eqref{eq_exchange}-\eqref{eq_exchange_2} do not modify the kinetic-energy terms in Eq.~\eqref{NLS}. For the sake of presentation, we henceforth set $\gamma\!=\!0$ (except otherwise stated), but we do keep in mind that these terms can be readily added in the description without affecting the results~\cite{footnote}.

\section{A quantum many-body approach}\label{section_quantum}

Our approach consists in three successive steps [Fig.~\ref{fig_approach}]:
\begin{itemize}
\item We introduce a parent quantum many-body Hamiltonian, whose semiclassical dynamics reproduces the time evolution of the driven nonlinear system in Eq.~\eqref{NLS_time}; 
\item Within this quantum framework, we derive the effective (Floquet) Hamiltonian that well captures the long time dynamics in the high-frequency limit ($2\pi/T\rightarrow\infty$); \item We then obtain the effective classical equations of motion from the effective quantum Hamiltonian. 
\end{itemize}

The validity of this approach will then be verified in Section~\ref{section_numerics}, through numerical studies of both quantum and classical dynamics.

\subsection{The parent quantum many-body system}

\begin{figure}[h!]
\includegraphics[width = \linewidth]{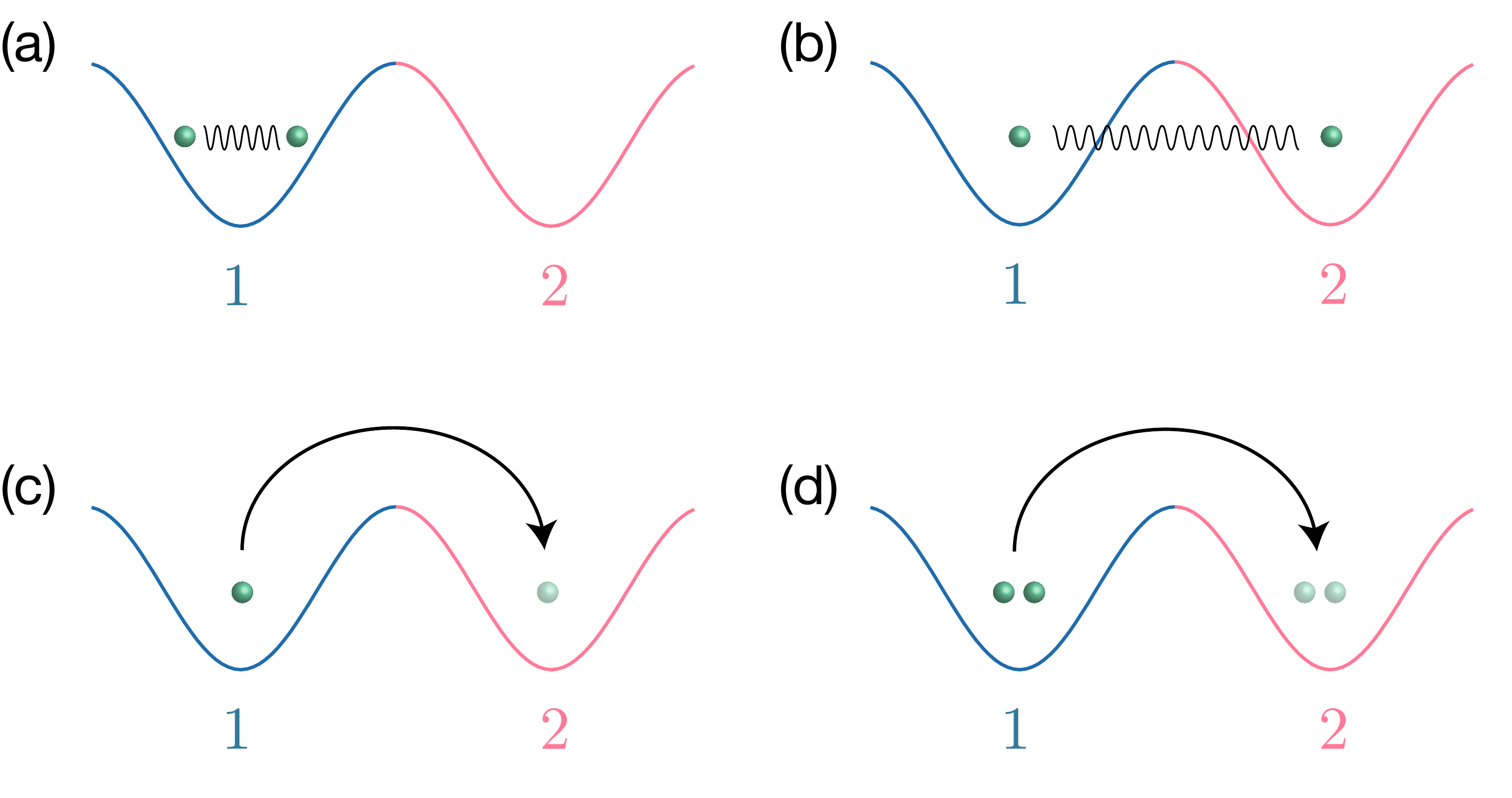}
\caption{Processes in the Hamiltonian in Eq.~\eqref{eq_parent_static}:~(a) Intra-mode (Hubbard) interactions; (b) inter-mode (cross) interactions; and (c) single-particle hopping processes.~(d) The effective Hamiltonian in Eq.~\eqref{eff_bosonic} includes pair-hopping processes, by which two interacting particles in the same mode simultaneously change mode. In this illustration, the two modes $1$ and $2$ correspond to the low-energy orbitals of a double-well potential, and the bosons are represented by green atoms.
}
\label{fig_process}
\end{figure}

Our starting point is the quantum many-body Hamiltonian
\begin{align}
\hat H_0 =& \frac{1}{2} \left ( \hat a_1^{\dagger} \hat a_1^{\dagger} \hat a_1  \hat a_1 + \hat a_2^{\dagger} \hat a_2^{\dagger} \hat a_2 \hat a_2 \right ) \notag \\
&+ \beta \hat a_1^{\dagger} \hat a_2^{\dagger} \hat a_1  \hat a_2 - \frac{\Omega_0}{2} \left (\hat a_1^{\dagger} \hat a_2 + \hat a_2^{\dagger} \hat a_1  \right),\label{eq_parent_static}
\end{align}
where $\hat a_{\sigma}^{\dagger}$ (resp.~$\hat a_{\sigma}$) creates (resp. annihilates) a boson in the mode $\sigma\!=\!1,2$. These operators satisfy the bosonic commutation relations, $[\hat a_{\sigma},\hat a_{\sigma'}^{\dagger}]=\delta_{\sigma,\sigma'}$. The first line in Eq.~\eqref{eq_parent_static} describes intra-mode (Hubbard) interactions, while the second line describes inter-mode (cross) interactions of strength $\beta$; the Hamiltonian also includes single-particle hopping processes of amplitude $\Omega_0/2$; see Fig.~\ref{fig_process}(a)-(c) for a sketch of the processes and Refs.~\cite{bloch2008many,dutta2015non}. Henceforth, the Hubbard interaction strength $U\!=\!1$ sets our unit of energy, as well as our unit of time $t_\text{unit}\!=\!\hbar/U$.

First of all, we note that the classical equations of motion in Eq.~\eqref{NLS} are readily obtained from Heisenberg's equations, $d \hat a_{\sigma}/dt = i [\hat H_0 , \hat a_{\sigma}]$, upon taking the classical limit $\hat a_{1,2} \rightarrow \psi_{1,2}$; see Refs.~\cite{holthaus2001towards,carusotto2013quantum,cao2020reconfigurable}. Specifically, the self-phase modulation in Eq.~\eqref{NLS} stems from the intra-mode (Hubbard) interaction terms in Eq.~\eqref{eq_parent_static}, while the cross-phase modulation stems from the inter-mode (cross) interaction term. Hence, this justifies the choice of Eq.~\eqref{eq_parent_static} as a proper parent quantum Hamiltonian for our initial (non-driven) system. Note that we set $\hbar\!=\!1$ throughout this work.

In fact, for the sake of later convenience, it is instructive to derive the classical equations of motion in Eq.~\eqref{NLS} using a different approach. Indeed, this will allow us to introduce central notions and quantities, which will be used throughout this work. Let us introduce a set of angular momentum (Schwinger) operators~\cite{auerbach2012interacting}, defined as
\begin{align}
&\hat J_x = \frac{1}{2} \left (\hat a_1^{\dagger} \hat a_2 + \hat a_2^{\dagger} \hat a_1 \right ) , \quad \hat J_y = \frac{1}{2i} \left (\hat a_2^{\dagger} \hat a_1 - \hat a_1^{\dagger} \hat a_2 \right ), \notag \\
&\hat J_z = \frac{1}{2} \left (\hat a_2^{\dagger} \hat a_2 - \hat a_1^{\dagger} \hat a_1 \right ) , \quad \hat N = \hat a_1^{\dagger} \hat a_1 + \hat a_2^{\dagger} \hat a_2. \label{Schwinger}
\end{align}
These operators satisfy the spin commutation relations $[\hat J_{\mu}, \hat J_{\nu}]=i\varepsilon_{\mu \nu \lambda} \hat J_{\lambda}$, and the operator $\hat N$ counts the total number of bosons in the system (assumed to be constant); note that $\hat J_{\mu}\!=\!\hat \sigma_{\mu}/2$ for a single boson ($N\!=\!1$), where $\hat \sigma_{x,y,z}$ denote the Pauli matrices. Using the operators in Eq.~\eqref{Schwinger}, the parent Hamiltonian in Eq.~\eqref{eq_parent_static} simply reads
\begin{equation}
\hat H_0 = \chi \hat J_z^2 - \Omega_0 \hat J_x + \text{constant}, \qquad \chi\!=\!1 - \beta,\label{Josephson}
\end{equation}
and we henceforth neglect the constant terms (proportional to $\hat N$ and $\hat N^2$); see Appendix~\ref{sect_app}. We note that the Hamiltonian in Eq.~\eqref{Josephson} has been extensively studied in the context of the bosonic Josephson effect~\cite{zibold2010classical,zibold2012classical} and  nuclear physics~\cite{lipkin1965validity}.


The equations of motion associated with Eq.~\eqref{Josephson} are readily obtained from Heisenberg's equations
\begin{align}
&\frac{d \hat J_z (t)}{dt} = i [ \hat H_0, \hat J_z (t) ] = - \Omega_0 \hat J_y (t), \label{eq_spin} \\
&\frac{d \hat J_y (t)}{dt} = i [ \hat H_0, \hat J_y (t) ] = \Omega_0 \hat J_z (t) \notag \\
&\hspace{3.5cm}+ \chi \left (\hat J_z (t)\hat J_x (t) + \hat J_x (t)\hat J_z (t) \right ). \notag
\end{align}

In order to connect Eq.~\eqref{eq_spin} to the classical nonlinear Schr\"odinger equation in Eq.~\eqref{NLS}, we take the classical limit and introduce the Bloch-Poincar\'e sphere representation $(\theta,\varphi)$ through the mapping
\begin{align}
&\hat J_x \rightarrow \frac{N}{2}\sqrt{1-z^2} \cos \varphi , \quad \hat J_y \rightarrow -\frac{N}{2}\sqrt{1-z^2} \sin \varphi ,\notag\\
&\hat J_z \rightarrow - \frac{N}{2} z, \hspace{2cm} z=\cos \theta . \label{mapping}
\end{align}
Injecting this into Eq.~\eqref{eq_spin}, one obtains the classical equations of motion
\begin{align}
&\dot z=- \Omega_0 \sqrt{1-z^2} \sin \varphi , \notag \\
&\dot \varphi= N \chi z + \Omega_0 \frac{z}{\sqrt{1-z^2}} \cos \varphi , \label{pendulum}
\end{align}
for the two canonical conjugate variables $z(t)$ and $\varphi(t)$~\cite{zibold2010classical,smerzi1997quantum}. 

We point out that Eq.~\eqref{pendulum} is equivalent to the nonlinear Schr\"odinger equation in Eq.~\eqref{NLS} upon representing the complex amplitudes $\psi_{1,2}$ on the Bloch-Poincar\'e sphere~\cite{paraoanu2001josephson,cao2020reconfigurable}
\begin{align}
& \psi_1=\sqrt{N}\cos (\theta/2)\, e^{-i \varphi/2} = \sqrt{\frac{N}{2}+n} \, e^{-i \varphi/2}, \notag \\
& \psi_2=\sqrt{N}\sin (\theta/2)\, e^{i \varphi/2} =\sqrt{\frac{N}{2}-n} \, e^{i \varphi/2},\label{psi_z_phi}
\end{align}
where we introduced the relative phase $\varphi$ between the two modes, the relative population (or relative light intensity)
\begin{equation}
z=\cos \theta = \frac{2n}{N}=\left ( \vert \psi_1 \vert^2 - \vert \psi_2 \vert^2  \right)/N,\label{z_eq}
\end{equation}
and the total population (or total light intensity) 
\begin{equation}
N=\vert \psi_1 \vert^2 + \vert \psi_2 \vert^2. 
\end{equation}
We emphasize that the dynamics in phase space, i.e.~the trajectories ($z(t),\varphi(t))$, can be simply monitored in an optical setting by measuring the light intensity and the relative phase of the two modes.

For the sake of completeness, we note that the equations of motion in Eq.~\eqref{pendulum} can be derived from Hamilton's equation, using the classical Hamiltonian~\cite{zibold2010classical,smerzi1997quantum,di2019nonlinear}
\begin{equation}
\mathcal{H}_0 (z,\varphi)= \frac{\chi N}{2} z^2 - \Omega_0 \sqrt{1-z^2} \cos \varphi .\label{classical_ham}
\end{equation}
The classical dynamics of the non-driven system hence relies on a competition between the ``mean-field" interaction parameter $g\!=\!\chi N$ and the linear coupling $\Omega_0$. This competition is at the core of bifurcations and symmetry breaking in the bosonic Josephson effect~\cite{zibold2010classical,cao2017experimental,cao2020reconfigurable}.

\subsection{The pulse sequence and the effective Floquet Hamiltonian}\label{sect_effective}

We now introduce the quantum-many-body analogue of the pulse sequence introduced in Eqs.~\eqref{eq_exchange}-\eqref{pulse}. We write the time-evolution operator over one period $T$ in the form [Fig.~\ref{fig1}(b)]
\begin{equation}
\hat U (T;0) = \hat U_{\text{mix}}^{\dagger} \, 
e^{-i \frac{T}{2} \hat H_0} \hat U_{\text{mix}} \, e^{-i \frac{T}{2} \hat H_0},\label{U_sequence}
\end{equation}
where the mixing operator is defined as
\begin{equation}
\hat U_{\text{mix}}=e^{i \frac{\pi}{2} \hat J_x}.
\end{equation}
We note that this indeed corresponds to the $\pi/2$-pulse operator in Eq.~\eqref{pi_over_two} for a single boson ($N\!=\!1$), which is consistent with the fact that the mixing operation is a single-particle process. We also point out that we explicitly took the limit $\tau\!\rightarrow\!0$, where $\tau$ is the pulse duration; see Eq.~\eqref{pulse}.

The state of the quantum many-body system at time $t_{\frak{n}}\!=\!T\!\times\!\frak{n} $ is then obtained as
\begin{equation}
\vert \psi (t_{\frak{n}}) \rangle = \hat U (t_{\frak{n}}; 0) \vert \psi (0) \rangle = \left (\hat U (T;0) \right )^{\frak{n}}\vert \psi (0) \rangle ,\label{time_evolution}
\end{equation}
where $\vert \psi (0) \rangle$ denotes the initial state of the system.

We now derive the effective (Floquet) Hamiltonian~\cite{goldman2014periodically,goldman2015periodically,bukov2015universal}, which captures the stroboscopic dynamics of the driven system, and hence, its time evolution over long time scales $t_{\frak{n}}\!\gg\!T$. The effective Hamiltonian is defined through the time-evolution operator over one period~\cite{kitagawa2010topological,goldman2014periodically}
\begin{equation}
\hat U (T;0) = e^{-i T \hat H_{\text{eff}}},\label{effective_definition}
\end{equation}
and it can be evaluated explicitly through a $1/\omega$-expansion, where $\omega\!=\!2\pi/T$ denotes the drive frequency; see Refs.~\cite{goldman2014periodically,goldman2015periodically,bukov2015universal,eckardt2015high,mikami2016brillouin}. In order to reach convergence of this infinite series expansion, we partially resum the series~\cite{goldman2014periodically} by splitting the time-evolution operator in Eq.~\eqref{U_sequence} into two parts
\begin{equation}
\hat U (T;0) = e^{-i \frac{T}{2} \hat H_1} e^{-i \frac{T}{2} \hat H_0},\label{split_eq}
\end{equation}
where we introduced the operator $\hat H_1$ defined as
\begin{equation}
e^{-i \frac{T}{2} \hat H_1}  \equiv  e^{-i \frac{\pi}{2} \hat J_x} 
e^{-i \frac{T}{2} \hat H_0} e^{i \frac{\pi}{2} \hat J_x}.\label{H1_def}
\end{equation}
Then, assuming that $T \omega_{\text{eff}}\!\ll\!1$, where $\omega_{\text{eff}}$ is the characteristic frequency associated with the processes included in the Hamiltonians $\hat H_{0}$ and $\hat H_{1}$, we apply the Trotter approximation to Eq.~\eqref{split_eq},
\begin{equation}
\hat U (T;0) \approx e^{-i \frac{T}{2} (\hat H_0+ \hat H_1)},\label{Trotter}
\end{equation}
from which we directly obtain the effective Hamiltonian [Eq.~\eqref{effective_definition}]
\begin{equation}
\hat H_{\text{eff}}=\frac{1}{2} (\hat H_0+ \hat H_1) + \mathcal{O} (T).\label{eff_interm}
\end{equation}

Our problem of finding the effective Hamiltonian thus reduces to the calculation of $\hat H_1$ defined in Eq.~\eqref{H1_def}. This step can be performed exactly, by noting that
\begin{equation}
\hat H_1 = e^{-i \frac{\pi}{2} \hat J_x} \hat H_0 \, e^{i \frac{\pi}{2} \hat J_x}= \chi \left (e^{-i \frac{\pi}{2} \hat J_x} \hat J_z^2 \, e^{i \frac{\pi}{2} \hat J_x} \right ) - \Omega_0 \hat J_x ,
\end{equation}
where we used the definition of $\hat H_0$ in Eq.~\eqref{Josephson}. Using the Baker-Campbell-Hausdorff formula, one obtains~\cite{kidd2019quantum}
\begin{equation}
e^{-i \frac{\pi}{2} \hat J_x} \hat J_z^2 \, e^{i \frac{\pi}{2} \hat J_x} = \hat J_y^2,\label{BCH}
\end{equation}
such that 
\begin{equation}
\hat H_1 = \chi \hat J_y^2 - \Omega_0 \hat J_x.\label{H1}
\end{equation}
The effective Hamiltonian in Eq.~\eqref{eff_interm} finally reads
\begin{equation}
\hat H_{\text{eff}}=\frac{\chi}{2} (\hat J_y^2 + \hat J_z^2) - \Omega_0 \hat J_x + \mathcal{O} (T).\label{eff_final}
\end{equation}
From this result, we find that the Trotter approximation [Eq.~\eqref{Trotter}] is valid for a sufficiently short driving period satisfying $T\!\ll\!1/\chi$ and $T\!\ll\!1/\Omega_0$. 

It is instructive to rewrite the effective Hamiltonian in Eq.~\eqref{eff_final} using the original bosonic operators [Appendix~\ref{sect_app}],
\begin{align}
\hat H_{\text{eff}} &= \frac{\chi}{8} \left ( \hat a_1^{\dagger} \hat a_1^{\dagger} \hat a_1  \hat a_1 + \hat a_2^{\dagger} \hat a_2^{\dagger} \hat a_2 \hat a_2 \right ) \notag \\
&- \frac{\chi}{8} \left ( \hat a_1^{\dagger} \hat a_1^{\dagger} \hat a_2  \hat a_2 + \hat a_2^{\dagger} \hat a_2^{\dagger} \hat a_1 \hat a_1 \right )\notag \\
& - \frac{\Omega_0}{2} \left (\hat a_1^{\dagger} \hat a_2 + \hat a_2^{\dagger} \hat a_1  \right) + \mathcal{O} (T) .\label{eff_bosonic}
\end{align}
A comparison with the initial Hamiltonian $\hat H_0$ in Eq.~\eqref{eq_parent_static} indicates that the driving pulse sequence has effectively generated novel interaction terms; see the second line of Eq.~\eqref{eff_bosonic}. These ``pair-hopping" terms~\cite{dutta2015non,anisimovas2015role} describe processes by which two particles in mode $\sigma$ collide and end up in the other mode $\sigma'\!\ne\!\sigma$; see Fig.~\ref{fig_process}(d). As we now discuss below, these pair-hopping terms are at the origin of the four-wave mixing nonlinearity announced in Eq.~\eqref{NLS_effective}. We also note that the effective interaction strength is given by $U_{\text{eff}}=\chi/8\!=\!(1-\beta)/8$, where $\beta$ sets the strength of the inter-mode (cross) interaction in the initial Hamiltonian $\hat H_0$ in Eq.~\eqref{eq_parent_static}.

\subsection{Effective classical equations of motion}\label{sect_classical_eff}

First of all, we find that the effective nonlinear Schr\"odinger equation in Eq.~\eqref{NLS_effective} is directly obtained from the effective Hamiltonian $\hat H_{\text{eff}}$ in Eq.~\eqref{eff_bosonic}, using Heisenberg’s equations $d \hat a_{\sigma}/dt = i [\hat H_{\text{eff}} , \hat a_{\sigma}]$, and upon taking the classical limit $\hat a_{1,2} \rightarrow \psi_{1,2}$. In particular, the effective four-wave mixing in Eq.~\eqref{NLS_effective} originates from the effective pair-hopping terms in Eq.~\eqref{eff_bosonic}.

In analogy with Eqs.~\eqref{eq_spin}-\eqref{pendulum}, we explicitly derive the classical equations of motion for the two canonical conjugate variables $z(t)$ and $\varphi(t)$, describing the relative population and phase of the two modes. Using the effective Hamiltonian in Eq.~\eqref{eff_final} and Heisenberg’s equations, we find
\begin{align}
&\frac{d \hat J_z (t)}{dt} = i [ \hat H_{\text{eff}}, \hat J_z (t) ] = - \Omega_0 \hat J_y (t) \notag \\
&\hspace{3.5cm}-\frac{\chi}{2} \left (\hat J_y (t)\hat J_x (t) + \hat J_x (t)\hat J_y (t) \right ), \notag \\
&\frac{d \hat J_y (t)}{dt} = i [ \hat H_{\text{eff}}, \hat J_y (t) ] = \Omega_0 \hat J_z (t) \notag \\
&\hspace{3.5cm}+\frac{\chi}{2} \left (\hat J_z (t)\hat J_x (t) + \hat J_x (t)\hat J_z (t) \right ). \notag
\end{align}
Finally, applying the Bloch-Poincar\'e-sphere mapping in Eq.~\eqref{mapping}, we obtain the classical equations of motion
\begin{align}
&\dot z=- \frac{\chi N}{2} (1-z^2) \cos \varphi \sin \varphi - \Omega_0 \sqrt{1-z^2} \sin \varphi, \notag \\
&\dot \varphi= \frac{\chi N}{2} z  \cos^2 \varphi + \Omega_0 \frac{z}{\sqrt{1-z^2}} \cos \varphi.\label{pendulum_eff}
\end{align}
We find that the equations of motion~\eqref{pendulum_eff} can be derived from Hamilton's equation, using the classical Hamiltonian
\begin{equation}
\mathcal{H}_{\text{eff}}(z,\varphi)= \frac{\chi N}{4} \left (z^2\cos^2 \varphi +\sin^2 \varphi \right ) - \Omega_0 \sqrt{1-z^2} \cos \varphi .\label{classical_eff}
\end{equation}
We stress that the classical equations of motion in Eq.~\eqref{pendulum_eff} are physically equivalent to the effective nonlinear Schr\"odinger equation announced in Eq.~\eqref{NLS_effective}, through the mapping provided by Eq.~\eqref{psi_z_phi}.

\section{Numerical analysis}\label{section_numerics}

This Section aims at exploring the validity of the effective-Hamiltonian analysis developed in Section~\ref{sect_effective} and its classical limit presented in Section~\ref{sect_classical_eff}.

\subsection{Validating the effective quantum Hamiltonian}

First, we demonstrate that the dynamics associated with the effective Hamiltonian in Eq.~\eqref{eff_bosonic} reproduces the stroboscopic dynamics of the driven system described by Eqs.~\eqref{U_sequence}-\eqref{time_evolution}. To this end, we choose a coherent spin state as an initial state~\cite{kitagawa1993squeezed,zibold2012classical}
\begin{equation}
\vert \psi (0) \rangle = \vert N , \theta, \varphi \rangle = \left (\hat a_{\theta,\varphi}^{\dagger} \right )^{N} \vert \emptyset \rangle,\label{CSS}
\end{equation}
which corresponds to a macroscopic occupation of the single-particle state,
\begin{equation}
\vert \theta, \varphi \rangle = \cos (\theta/2) \vert 1 \rangle + \sin(\theta/2)e^{i \varphi} \vert 2 \rangle ,
\end{equation}
defined on the Bloch sphere. Here we introduced the single-particle states $\vert 1 \rangle\!=\!\hat a_{1}^{\dagger} \vert \emptyset \rangle$ and $\vert 2 \rangle\!=\!\hat a_{2}^{\dagger} \vert \emptyset \rangle$, associated with the two modes, as well as the creation operator $\hat a_{\theta,\varphi}^{\dagger}\vert \emptyset \rangle\!=\!\vert \theta, \varphi \rangle$. We note that the chosen initial state in Eq.~\eqref{CSS} behaves classically in the limit $N\!\rightarrow\!\infty$~\cite{zibold2012classical}, which will be convenient for later purposes (i.e.~when comparing quantum and classical dynamics). 

\begin{figure}[h!]
\includegraphics[width = \linewidth]{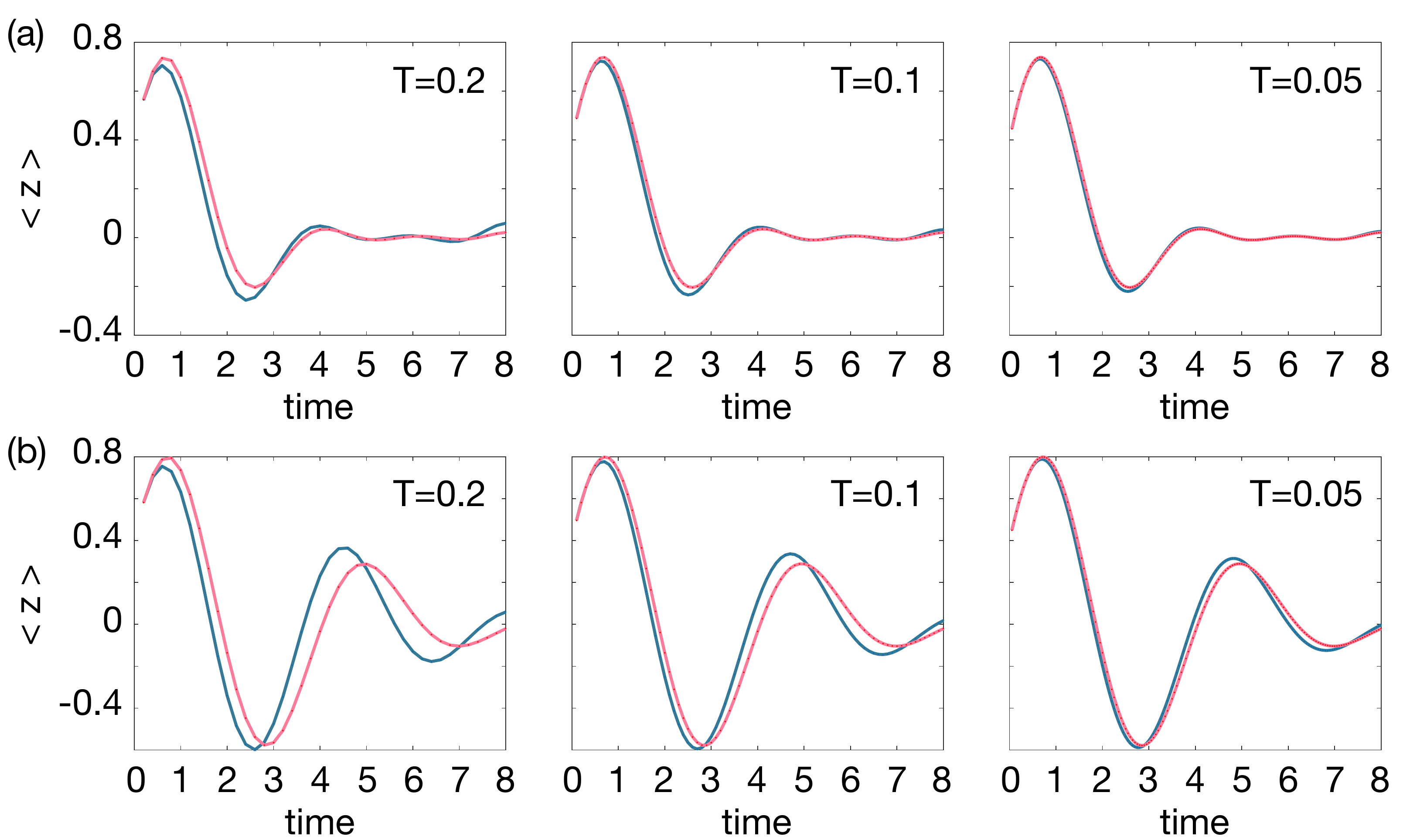}
\caption{Population imbalance $\langle z \rangle$ as a function of time, as obtained from the quantum dynamics of the driven system (blue curve) and the effective-Hamiltonian quantum dynamics (red curve) for: (a) $N\!=\!10$ bosons and (b) $N\!=\!50$ bosons. For each case, the full time dynamics of the driven system is generated using the sequence in Eq.~\eqref{U_sequence} with a period $T\!=\!0.2$, $T\!=\!0.1$ and $T\!=\!0.05$. Here the interaction parameter is set to $g\!=\!\chi N\!=\!5$ and the static linear coupling is set to $\Omega_0\!=\!0$; the initial coherent spin state $\vert N , \theta, \varphi \rangle$ corresponds to $z\!=\!\cos \theta\!=\!0.4$ and $\varphi\!=\!2.25$. In all plots, the time-evolved state is evaluated at stroboscopic times $t_{\frak{n}}\!=\!T\!\times\!\frak{n} $.
}
\label{fig_effective_full_quantum}
\end{figure}

We analyze the quantum dynamics through the evaluation of the population imbalance 
\begin{equation}
\langle z (t_{\frak{n}}) \rangle = (2/N) \langle \psi (t_{\frak{n}}) \vert \hat J_z \vert \psi (t_{\frak{n}}) \rangle , \quad t_{\frak{n}}\!=\!T\!\times\!\frak{n} , \notag
\end{equation}
where the time-evolved state $\vert \psi (t_{\frak{n}}) \rangle$ is obtained from:~(i) the full time dynamics of the driven system [Eqs.~\eqref{U_sequence}-\eqref{time_evolution}], and (ii) the effective Hamiltonian [Eq.~\eqref{eff_bosonic}]. Figure~\ref{fig_effective_full_quantum} compares these two results for both $N\!=\!10$ and $N\!=\!50$ bosons, and the same interaction parameter $g\!=\!\chi N\!=\!5$. In both cases, one obtains that the effective description well captures the stroboscopic dynamics when the driving period is sufficiently small, $T\!\lesssim\!0.1$ in the current units [see Eq.~\eqref{eq_parent_static}]. This analysis validates the effective Hamiltonian in Eq.~\eqref{eff_bosonic} in the high-frequency regime.

\subsection{The effective semiclassical dynamics}

As a next step, we now show that the effective Hamiltonian $\hat H_{\text{eff}}$ in Eq.~\eqref{eff_bosonic} well captures the classical dynamics generated by the equations of motion in Eq.~\eqref{pendulum_eff}. We remind that the latter classical description is associated with the  Hamiltonian function $\mathcal{H}_{\text{eff}}(z,\varphi)$ displayed in Eq.~\eqref{classical_eff}, where $z$ and $\varphi$ describe the relative population and phase of the two modes; see Eqs.~\eqref{psi_z_phi}-\eqref{z_eq}. The agreement between the quantum and classical descriptions is expected to be reached in the limit $N\!\rightarrow\infty$, where quantum fluctuations become negligible~\cite{smerzi1997quantum,gajda1997fluctuations,zibold2010classical,zibold2012classical,carusotto2013quantum,larre2015propagation}. We also remind the reader that the classical equations of motion in Eq.~\eqref{pendulum_eff}, which are analyzed in this Section, are equivalent to the effective nonlinear Schr\"odinger equation in Eq.~\eqref{NLS_effective}, through the mapping defined in Eq.~\eqref{psi_z_phi}.

\begin{figure}[h!]
\includegraphics[width = \linewidth]{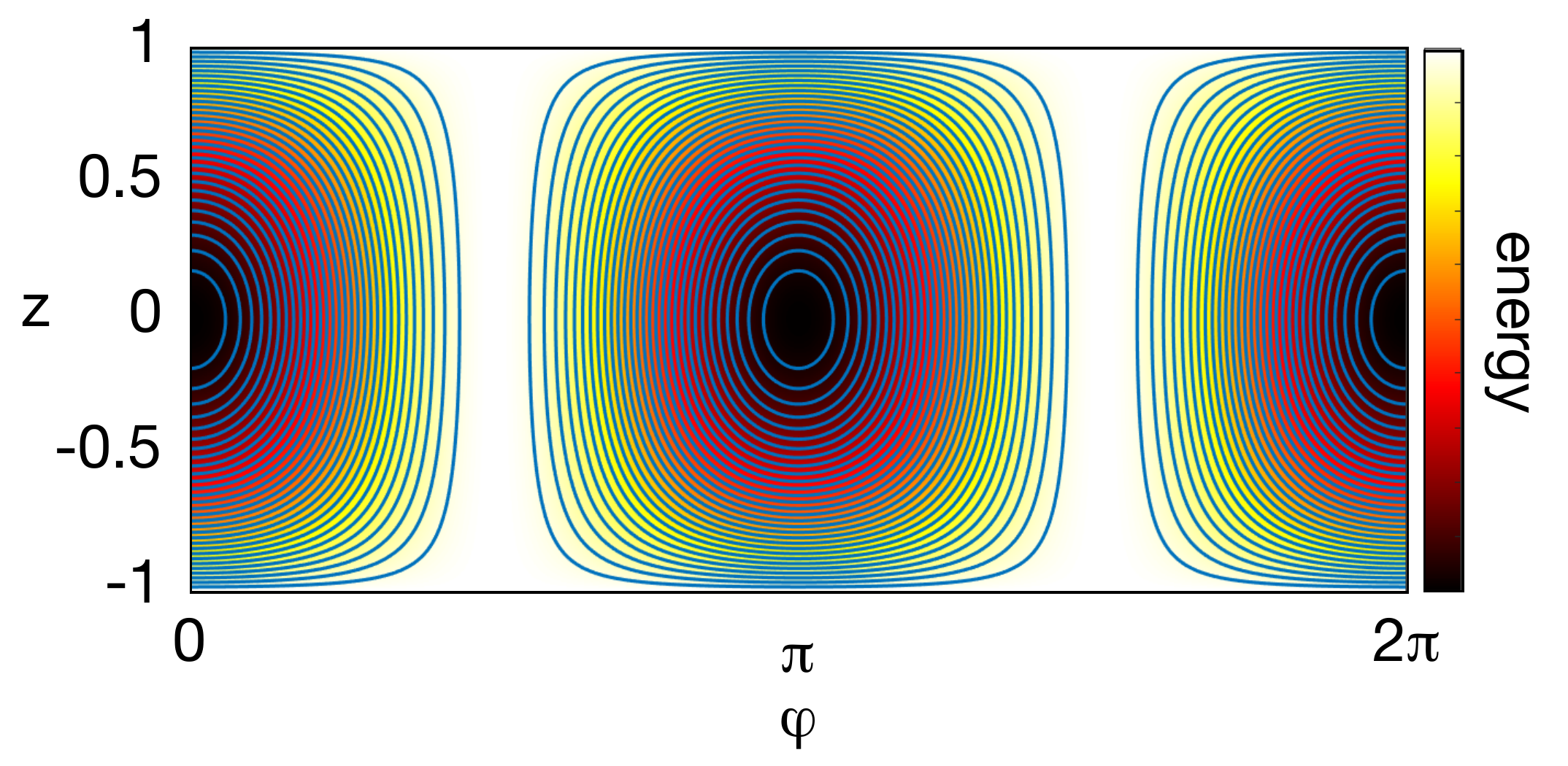}
\caption{Energy landscape associated with the classical Hamiltonian $\mathcal{H}_{\text{eff}}(z,\varphi)$ displayed in Eq.~\eqref{classical_eff}, for $\Omega_0\!=\!0$. A few trajectories are indicated as thin blue curves (equipotential lines of the energy landscape).
}
\label{fig_eff_landscape}
\end{figure}

First of all, let us analyze the dynamics generated by the effective classical equations of motion in Eq.~\eqref{pendulum_eff}. In order to highlight the role of nonlinearities, we hereby set the static linear coupling to $\Omega_0\!=\!0$. In Fig.~\ref{fig_eff_landscape}, we display a few representative trajectories over the energy landscape $\mathcal{H}_{\text{eff}}(z,\varphi)$ defined in Eq.~\eqref{classical_eff}. These trajectories reflect the presence of two stable fixed points at $(z\!=\!0,\varphi\!=\!0)$ and $(z\!=\!0,\varphi\!=\!\pi)$. We stress that this configuration of fixed points radically differs from that associated with the non-driven system [see $\mathcal{H}_0(z,\varphi)$ in Eq.~\eqref{classical_ham}] for the same choice of $\Omega_0\!=\!0$.

\begin{figure}[h!]
\includegraphics[width = \linewidth]{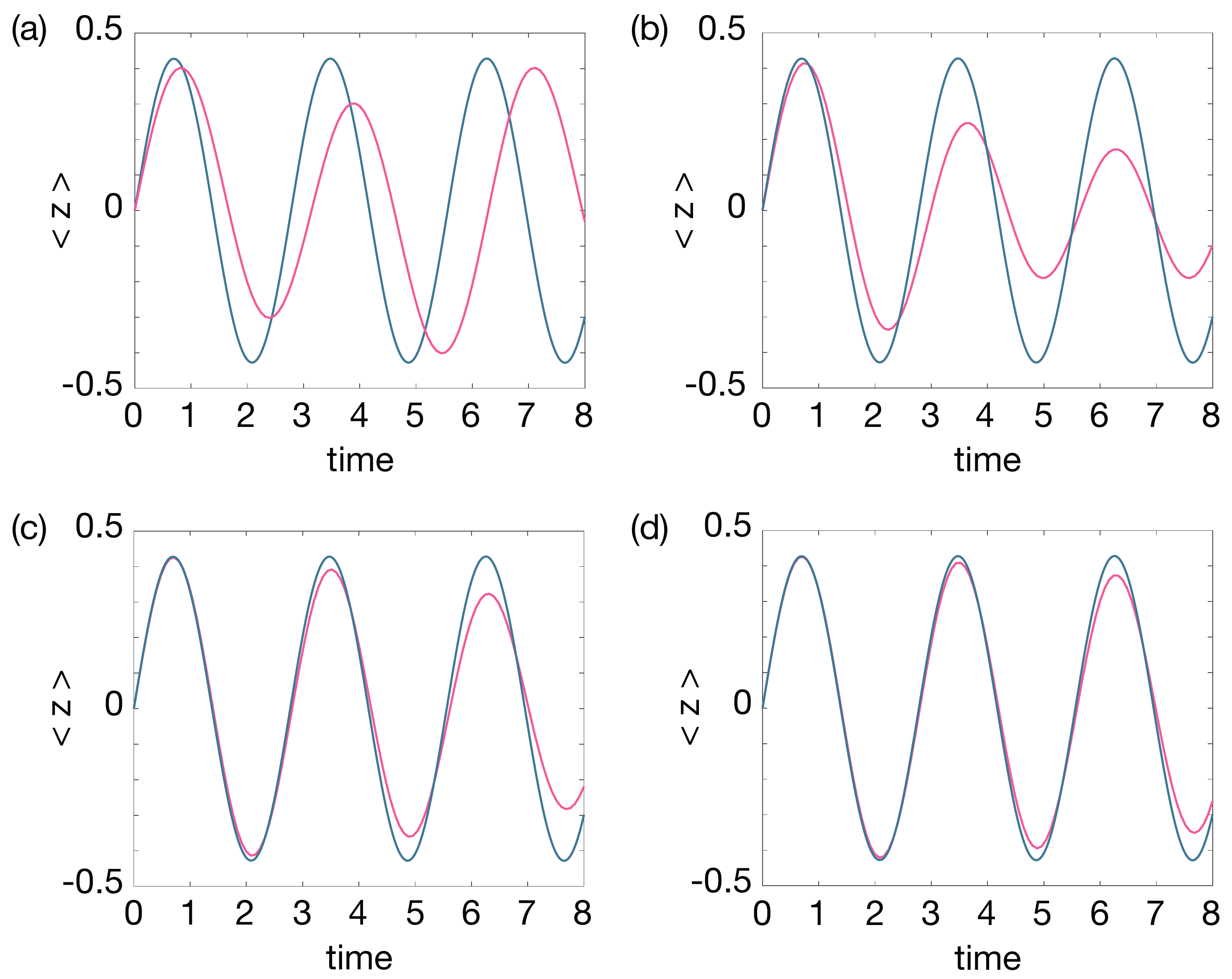}
\caption{Population imbalance $\langle z \rangle$ as a function of time, as obtained from the effective-Hamiltonian quantum dynamics (red curve) and the effective classical equations of motion (blue curve). The number of bosons is: (a) $N\!=\!5$; (b) $N\!=\!10$; (c) $N\!=\!80$; (d) $N\!=\!170$. Here the interaction parameter is set to $g\!=\!\chi N\!=\!5$, while the static linear coupling is set to $\Omega_0\!=\!0$; the initial coherent spin state $\vert N , \theta, \varphi \rangle$ corresponds to $z\!=\!\cos \theta\!=\!0$ and $\varphi\!=\!2.7$; the same initial condition is chosen for the effective classical dynamics.
}
\label{fig_eff_classical}
\end{figure}

We now compare these classical predictions to the quantum dynamics associated with the effective Hamiltonian $\hat H_{\text{eff}}$ in Eq.~\eqref{eff_bosonic}, using a coherent spin state $\vert N , \theta, \varphi \rangle$ as an initial condition; see Eq.~\eqref{CSS}. Figure~\ref{fig_eff_classical} shows the trajectories $\langle z(t) \rangle$ for $N\!=\!5, 10, 80, 170$ bosons, while keeping the ``mean-field" interaction parameter $\chi N\!=\!5$ constant. From these results, we confirm that a good agreement between the effective classical and quantum descriptions is indeed obtained in the large $N$ limit. 

\begin{figure}
\includegraphics[width=\linewidth]{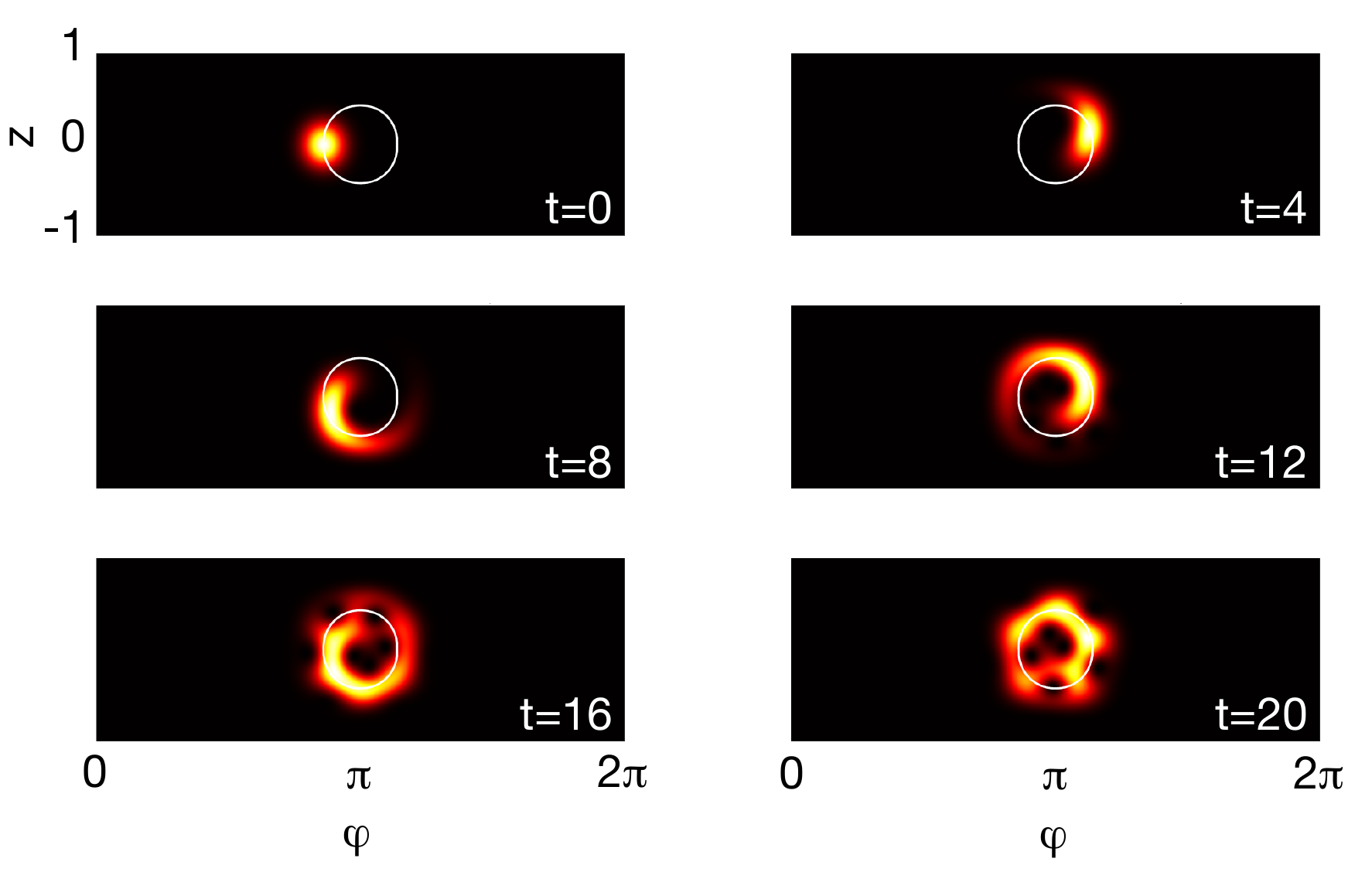}
\caption{Time-evolving Husimi function $Q(z,\varphi; t)$ for a state $\vert \psi (t) \rangle$ that evolves according to the effective Hamiltonian $\hat H_{\text{eff}}$ in Eq.~\eqref{eff_bosonic}. Here, the number of bosons is $N\!=\!80$, and the other parameters are the same as in Fig.~\ref{fig_eff_classical}(c). The initial coherent spin state $\vert \psi (0) \rangle\!=\!\vert N , \theta, \varphi \rangle$, at $z\!=\!\cos \theta\!=\!0$ and $\varphi\!=\!2.7$, becomes substantially squeezed around time $t\!\approx\!3$, hence signaling the breakdown of its classical description. An oversqueezed state, exhibiting Majorana stars, appears around $t\!\approx\!12$. The trajectory predicted by the effective classical equations of motion [Eq.~\eqref{pendulum_eff}] is depicted in white.
}
\label{fig_husimi}
\end{figure}

In order to further appreciate the residual deviations between the quantum and classical dynamics in the small $N$ regime, we depict the time-evolving Husimi function $Q(z,\varphi; t)$ in Fig.~\ref{fig_husimi} for the case $N\!=\!80$. The Husimi function~\cite{zibold2010classical,zibold2012classical,julia2012dynamic,bruno2012quantum,strobel2014fisher,evrard2019enhanced,nascimbene2020quantum} is obtained by evaluating the squared overlap of the time-evolving state $\vert \psi (t) \rangle$ with the coherent spin states defined over the Bloch sphere (with same particle number $N$),
\begin{equation}
Q(z,\varphi; t) = \vert \langle N , \theta, \varphi \vert \psi (t) \rangle \vert^2 , \quad z=\cos \theta .
\end{equation}
Here the state $\vert \psi (t) \rangle$ is evolved according to the effective Hamiltonian $\hat H_{\text{eff}}$ in Eq.~\eqref{eff_bosonic}, so that the evolution of the Husimi function in Fig.~\ref{fig_husimi} is to be compared with the quantum dynamics displayed in Fig.~\ref{fig_eff_classical}(c) for $N\!=\!80$ bosons. The time-evolution of the Husimi function $Q(z,\varphi; t)$ shown in Fig.~\ref{fig_husimi} indicates that the initial coherent spin state $\vert \psi (0) \rangle\!=\!\vert N , \theta, \varphi \rangle$ becomes substantially squeezed~\cite{kitagawa1993squeezed} around time $t\!\approx3$, which also corresponds to the time around which the classical trajectory starts deviating from the effective-Hamiltonian quantum dynamics in Fig.~\ref{fig_eff_classical}(c). At later times, $t\!\approx\!12$, the state becomes oversqueezed and it exhibits Majorana stars in the Husimi distribution~\cite{bruno2012quantum,evrard2019enhanced,nascimbene2020quantum}. We find that these non-classical features are postponed to later evolution times upon increasing the number of bosons $N$ while keeping the interaction parameter $g\!=\!\chi N$ fixed. Despite these non-classical features, the center of mass of the Husimi function is found to approximately follow a classical orbit around the stable fixed point $(z\!=\!0,\varphi\!=\!\pi)$, as depicted in Fig.~\ref{fig_eff_landscape}.

\subsection{The driven nonlinear Schr\"odinger equation \\ and its effective description}

In this Section, we analyze the agreement between the classical dynamics associated with the driven nonlinear Schr\"odinger equation [Eqs.~\eqref{NLS_time}-\eqref{pulse}] and the dynamics generated by the effective classical equations of motion [Eq.~\eqref{pendulum_eff}], which derive from the Hamiltonian $\mathcal{H}_{\text{eff}}(z,\varphi)$ in Eq.~\eqref{classical_eff}. We remind that these effective equations of motion are equivalent to the effective nonlinear Schr\"odinger equation announced in Eq.~\eqref{NLS_effective}.

In practice, we numerically solve the classical equations of motion [Eq.~\eqref{pendulum}]
\begin{align}
&\dot z=f_{\text{pulse}}(t) \sqrt{1-z^2} \sin \varphi , \notag \\
&\dot \varphi= N \chi z - f_{\text{pulse}}(t) \frac{z}{\sqrt{1-z^2}} \cos \varphi , \label{pendulum_driven}
\end{align}
where the pulse function $f_{\text{pulse}}(t)$ is defined in Eq.~\eqref{pulse}. These equations of motion are equivalent to the driven nonlinear Schr\"odinger equation in Eqs.~\eqref{NLS_time}-\eqref{pulse} through the mapping provided by Eq.~\eqref{psi_z_phi}.

\begin{figure}[h!]
\includegraphics[width = \linewidth]{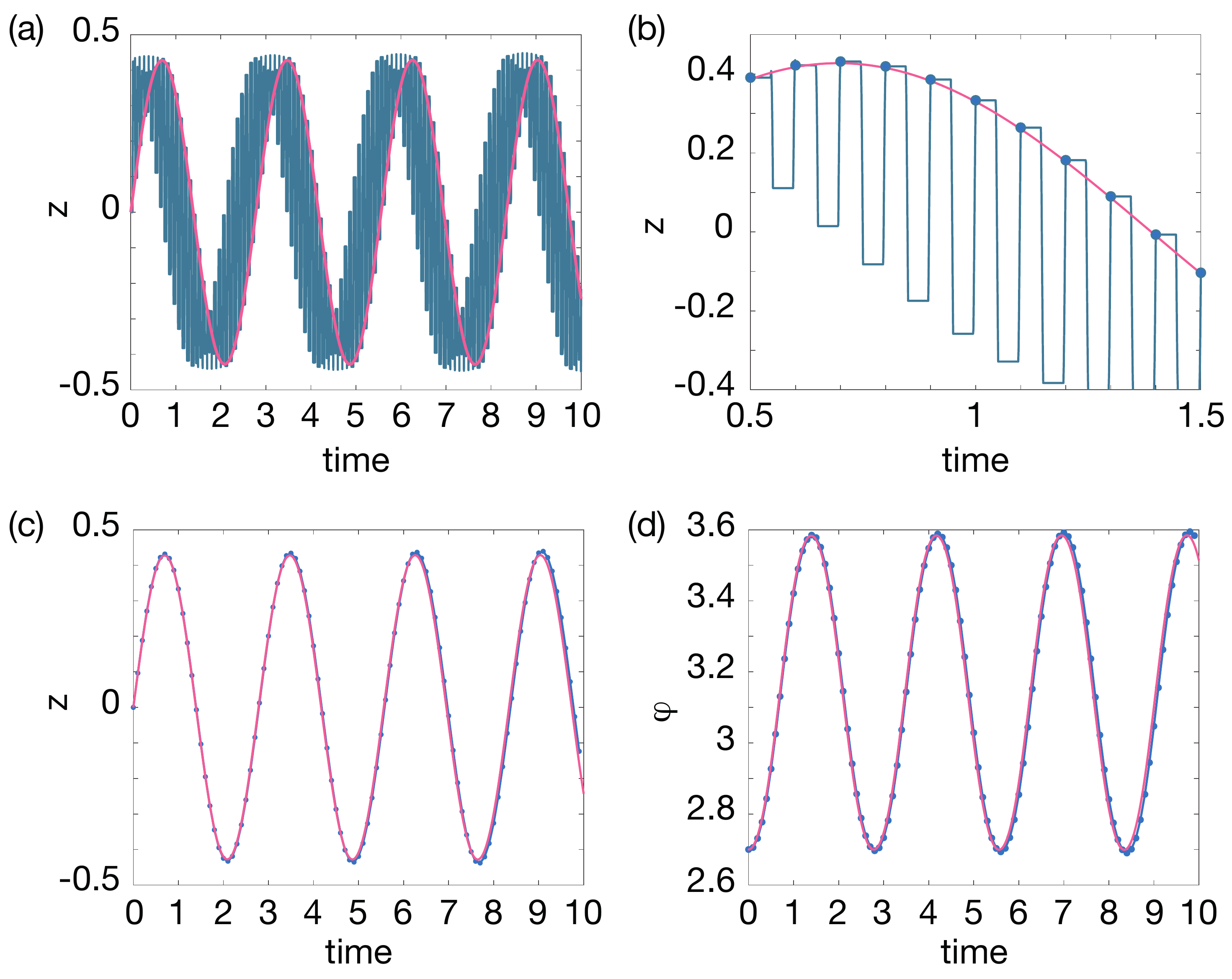}
\caption{The driven nonlinear Schr\"odinger equation versus its effective description:~(a) Population imbalance $z(t)$ as a function of time, as obtained from the driven nonlinear Schr\"odinger equation in Eq.~\eqref{pendulum_driven} (blue curve) and from the effective classical equations of motion in Eq.~\eqref{pendulum_eff} (red curve). (b) Zoom in the panel (a):~the blue dots highlight the stroboscopic dynamics at times $t_{\frak{n}}$; note the micromotion at arbitrary times $t\!\ne\!t_{\frak{n}}$. (c) Stroboscopic dynamics $z(t_{\frak{n}})$ obtained from the driven nonlinear Schr\"odinger equation (blue curve and dots), compared with the effective classical description (red curve). (d) Same as in panel (c) but for the other canonical variable $\varphi$. In all panels, the period of the drive is set to $T\!=\!0.1$ and the pulse duration to $\tau\!=\!T/20$; the interaction parameter is set to $g\!=\!\chi N\!=\!5$, while the static linear coupling is set to $\Omega_0\!=\!0$; the initial condition corresponds to $z\!=\!\cos \theta\!=\!0$ and $\varphi\!=\!2.7$ as in Fig.~\ref{fig_eff_classical}. 
}
\label{fig_classical_complete}
\end{figure}

The resulting dynamics are displayed in Fig.~\ref{fig_classical_complete}, together with the dynamics generated from the effective classical Hamiltonian $\mathcal{H}_{\text{eff}}(z,\varphi)$ in Eq.~\eqref{classical_eff}. The results in Fig.~\ref{fig_classical_complete} confirm that the effective classical description very well captures the  dynamics of the driven nonlinear system at stroboscopic times $t\!=\!t_{\frak{n}}$, while a finite micromotion is observed at intermediate times $t\!\ne\!t_{\frak{n}}$.\\


Altogether, the numerical studies presented in this Section~\ref{section_numerics} validate the effective description announced in Eq.~\eqref{NLS_effective} [see also Sections~\ref{sect_effective} and \ref{sect_classical_eff}], and hence, confirm the creation of effective interactions and nonlinearities through the repeated pulse sequence.

\section{Tuning interaction processes and classical nonlinearities}\label{sect_imbalanced}

\subsection{The imbalanced pulse sequence}

The pulse sequence introduced in Eq.~\eqref{U_sequence} corresponds to a balanced four-step sequence, with a free-evolution duration set to $T/2$ during the first and third steps of the sequence [Fig.~\ref{fig1}(b)]. However, it is instructive to consider the ``imbalanced" sequence
\begin{equation}
\hat U_{\alpha} (T;0) = \hat U_{\text{mix}}^{\dagger} \, 
e^{-i (1- \alpha) T \hat H_0} \hat U_{\text{mix}} \, e^{-i \alpha T \hat H_0},\label{U_sequence_imbalanced}
\end{equation}
where the parameter $\alpha$ quantifies the imbalance; note that $\alpha\!=\!1/2$ for the balanced sequence in Eq.~\eqref{U_sequence}. Following the approach of Section~\ref{sect_effective}, the effective Hamiltonian in Eqs.~\eqref{eff_interm} and \eqref{eff_final} is then generalized to
\begin{align}
\hat H_{\text{eff}}^{(\alpha)}&=\alpha \hat H_0 + (1-\alpha)\hat H_1 + \mathcal{O} (T) \label{effective_J_gen}\\
&= \chi \left (\alpha \hat J_z^2 + (1-\alpha) \hat J_y^2 \right ) - \Omega_0 \hat J_x + \mathcal{O} (T) . \notag
\end{align}
At this stage, it is important to consider two limiting cases:~when $\alpha\!=\!1$, one finds $\hat H_{\text{eff}}\!=\!\hat H_0$, which reflects the triviality of the sequence in Eq.~\eqref{U_sequence_imbalanced} in this case. When $\alpha\!=\!0$, one finds the effective Hamiltonian
\begin{align}
\hat H_{\text{eff}}^{(0)} = \chi \hat J_y^2 - \Omega_0 \hat J_x = e^{-i \frac{\pi}{2} \hat J_x} (\hat H_0) \, e^{i \frac{\pi}{2} \hat J_x} ,
\end{align}
which is thus strictly equivalent to the non-driven Hamiltonian $\hat H_0$ up to a unitary transformation [see Eq.~\eqref{BCH}]:~the Hamiltonians $\hat H_0$ and $\hat H_{\text{eff}}^{(0)}$ share the same spectrum. In fact, in the ``pathological" case $\alpha\!=\!0$, the driving sequence simply generates an initial and final kick~\cite{goldman2014periodically}, as can be deduced by explicitly writing the time-evolving state at some arbitrary stroboscopic time $t\!=\!t_{\frak{n}}$ [Eq.~\eqref{time_evolution}]
\begin{align}
\vert \psi (t_{\frak{n}}) \rangle &= \left (\hat U_{\alpha=0} (T;0) \right )^{\frak{n}}\vert \psi (0) \rangle ,\notag \\
&=e^{-i \frac{\pi}{2} \hat J_x} e^{-i t_{\frak{n}} \hat H_0} e^{i \frac{\pi}{2} \hat J_x} \vert \psi (0) \rangle.\label{time_evolution_kick}
\end{align}
The long-time dynamics in Eq.~\eqref{time_evolution_kick} is indeed dictated by the static Hamiltonian $\hat H_0$, but it is also affected by the initial kick $e^{i \frac{\pi}{2} \hat J_x}$ and the final kick $e^{- i \frac{\pi}{2} \hat J_x}$. 

Altogether, one finds that non-trivial interaction (or nonlinearity) effects are generated by driving sequences corresponding to $\hat U_{\alpha} (T;0)$ in Eq.~\eqref{U_sequence_imbalanced} with $\alpha\!\ne\!0$ and $\alpha\!\ne\!1$. One indeed verifies that $\hat H_{\text{eff}}^{(\alpha)}$ in Eq.~\eqref{effective_J_gen} and $\hat H_0$ do not have the same spectrum in this case.


\subsection{The classical analysis: phase-space transitions and spontaneous symmetry breaking}\label{sect_ssb}

\begin{figure}[h!]
\includegraphics[width = \linewidth]{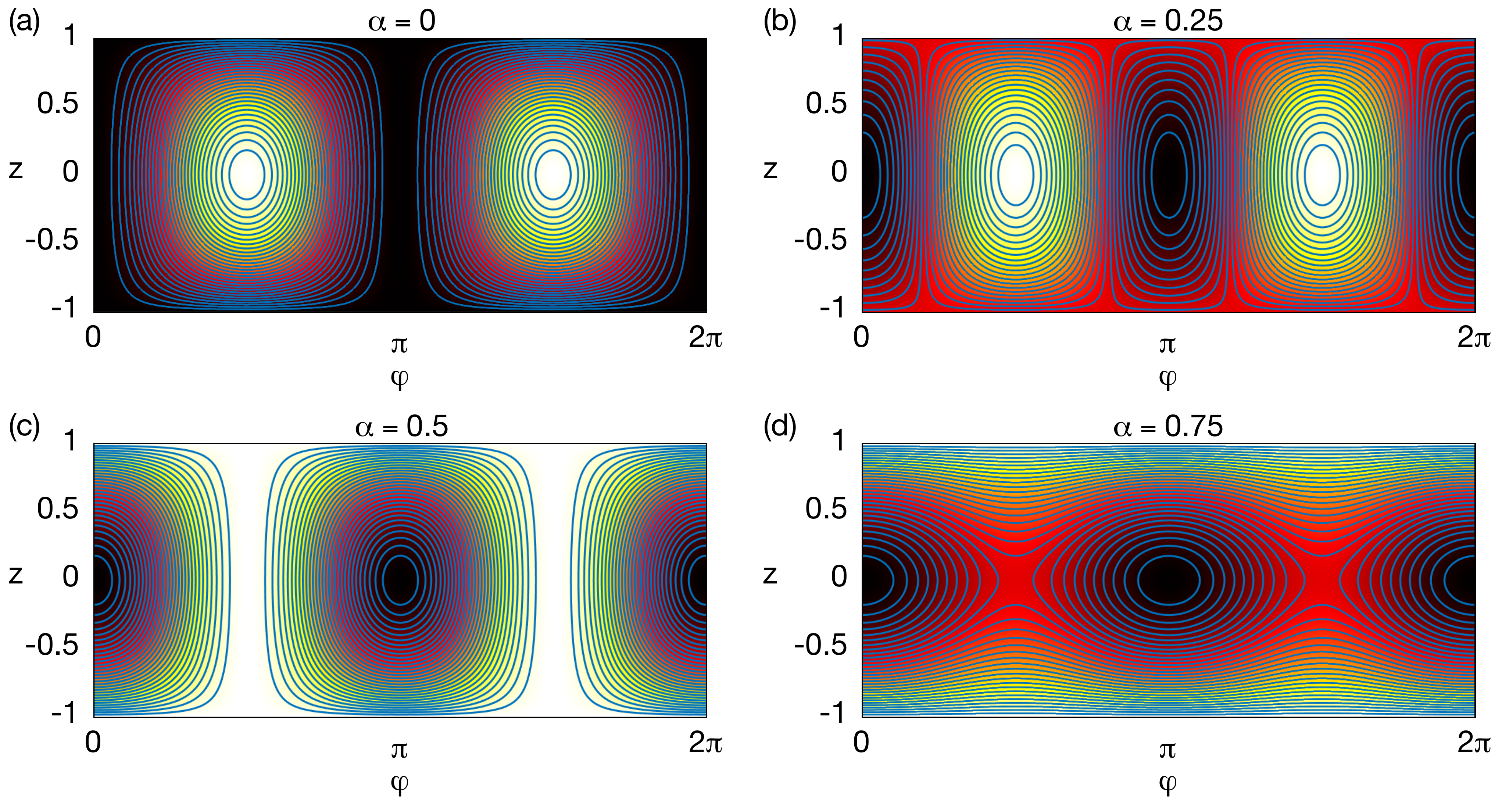}
\caption{Energy landscape associated with the effective classical Hamiltonian $\mathcal{H}_{\text{eff}}(z,\varphi;\alpha)$ displayed in Eq.~\eqref{classical_eff_imbalanced}, for four values of the imbalance parameter: (a) $\alpha\!=\!0$, (b) $\alpha\!=\!0.25$, (c) $\alpha\!=\!0.5$ and (d) $\alpha\!=\!0.5$. A few trajectories are indicated as thin blue curves (equipotential lines of the energy landscape) for each case. The static linear coupling is set to $\Omega_0\!=\!0$. Note the emergence and disappearance of stable fixed points on the Bloch-Poincar\'e sphere, as the imbalance parameter $\alpha$ is varied.
}
\label{fig_landscape_imbalanced}
\end{figure}

Following Section~\ref{sect_classical_eff}, we obtain the generalized classical equations of motion for the relative population and phase,
\begin{align}
&\dot z=- \chi N (1-\alpha) (1-z^2) \cos \varphi \sin \varphi - \Omega_0 \sqrt{1-z^2} \sin \varphi, \notag \\
&\dot \varphi= \chi N z \left (\alpha - (1-\alpha)  \sin^2 \varphi \right ) + \Omega_0 \frac{z}{\sqrt{1-z^2}} \cos \varphi. \notag
\end{align}
These equations of motion are found to derive from the classical Hamiltonian
\begin{align}
&\mathcal{H}_{\text{eff}}(z,\varphi;\alpha)=- \Omega_0 \sqrt{1-z^2} \cos \varphi  \label{classical_eff_imbalanced} \\
&\hspace{1.2cm}+\frac{\chi N}{2} \left (\alpha z^2 - (1-\alpha) z^2\sin^2 \varphi + (1-\alpha)\sin^2 \varphi \right ) .\notag
\end{align}
We represent the corresponding trajectories in Fig.~\ref{fig_landscape_imbalanced}, for various values of the imbalance parameter $\alpha$; in order to highlight the role of effective nonlinearities, we again set the static linear coupling to $\Omega_0\!=\!0$. Interestingly, the system undergoes a succession of transitions as the imbalance parameter $\alpha$ is varied, which are characterized by changes in the topology of phase space~\cite{zibold2010classical}:~When $\alpha\!=\!0$, the system is characterized by two stable classical fixed points at $(z\!=\!0, \varphi\!=\!\pi/2)$ and $(z\!=\!0, \varphi\!=\!3\pi/2)$ [Fig.~\ref{fig_landscape_imbalanced}(a)]; increasing $\alpha$ then generates two new stable fixed points at $(z\!=\!0, \varphi\!=\!0)$ and $(z\!=\!0, \varphi\!=\!\pi)$ [Fig.~\ref{fig_landscape_imbalanced}(b)]; the two initial fixed points at $(z\!=\!0, \varphi\!=\!\pi/2)$ and $(z\!=\!0, \varphi\!=\!3\pi/2)$ then become unstable at $\alpha\!=\!0.5$ [Fig.~\ref{fig_landscape_imbalanced}(c)], giving rise to two new stable fixed points located at the poles of the Bloch-Poincar\'e sphere $z\!=\!\pm 1$ [Fig.~\ref{fig_landscape_imbalanced}(d)]. We note that the emerging fixed points at $z\!=\!\pm 1$ are associated with the notion of spontaneous symmetry breaking, and were previously investigated in the context of ultracold gases~\cite{zibold2010classical,zibold2012classical} and in optical microcavities~\cite{cao2017experimental,cao2020reconfigurable}. In the present context, the symmetry breaking occurs as soon as the $\hat J_z^2$ interaction term dominates over the $\hat J_y^2$ interaction term; see the effective Hamiltonian $\hat H_{\text{eff}}^{(\alpha)}$ in Eq.~\eqref{effective_J_gen}. We finally point out that the fixed points at $(z\!=\!0, \varphi\!=\!0)$ and $(z\!=\!0, \varphi\!=\!\pi)$ become unstable when $\alpha\!=\!1$ (not shown in Fig.~\ref{fig_landscape_imbalanced}).

\subsection{Tunable interactions and nonlinearities}

We now rewrite the generalized effective Hamiltonian in Eq.~\eqref{effective_J_gen} in terms of the original bosonic operators [Eq.~\eqref{Schwinger}],
\begin{align}
\hat H_{\text{eff}}^{(\alpha)} &= \frac{U_1}{2} \left ( \hat a_1^{\dagger} \hat a_1^{\dagger} \hat a_1  \hat a_1 + \hat a_2^{\dagger} \hat a_2^{\dagger} \hat a_2 \hat a_2 \right ) \notag \\
&+U_2 \left ( \hat a_1^{\dagger} \hat a_2^{\dagger} \hat a_1  \hat a_2 \right )\notag \\
&+\frac{U_3}{2} \left ( \hat a_1^{\dagger} \hat a_1^{\dagger} \hat a_2  \hat a_2 + \hat a_2^{\dagger} \hat a_2^{\dagger} \hat a_1 \hat a_1 \right )\notag \\
& -\frac{\Omega_0}{2} \left (\hat a_1^{\dagger} \hat a_2 + \hat a_2^{\dagger} \hat a_1  \right) + \mathcal{O} (T) , \label{eff_bosonic_gen}
\end{align}
where the interaction strengths are given by
\begin{align}
&U_1= \alpha (1-\beta)/2, \notag \\
&U_2= (1-2 \alpha) (1- \beta)/2 , \notag \\
&U_3= (\alpha-1) (1-\beta)/2 . \label{g_def}
\end{align}
From this, we find that the imbalanced pulse sequence offers an efficient method to control the strength and sign of the different interaction processes.

This picture also offers an insightful view on the transition to spontaneous symmetry breaking discussed in Section~\ref{sect_ssb}:~this transition, which takes place at $\alpha\!=\!1/2$, results from a competition between the intra-mode (Hubbard) interaction strength $U_1$ and the pair-hopping strength $U_3$ in Eq.~\eqref{eff_bosonic_gen}, in the absence of linear coupling ($\Omega_0\!=\!0$). This is different from the transition discussed in Refs.~\cite{zibold2010classical,cao2017experimental}, which involves a competition between the Hubbard interaction strength $U_1$ and the linear coupling $\Omega_0$.

Finally, in the classical limit ($N\!\rightarrow\!\infty$), the effective nonlinear Schr\"odinger equation associated with the imbalanced pulse sequence reads
\begin{align}
&i \frac{\partial \psi_1}{\partial t} = \left ( - \gamma \frac{\partial^2}{\partial x^2} + U_1 \vert \psi_1 \vert^2 + U_2 \vert \psi_2 \vert^2    \right ) \psi_1 \notag \\
&\hspace{1.2cm} +U_3 \psi_1^* \psi_2^2 - \frac{\Omega_0}{2} \psi_2 , \notag\\
&i \frac{\partial \psi_2}{\partial t} = \left ( -\gamma \frac{\partial^2}{\partial x^2} + U_1 \vert \psi_2 \vert^2  + U_2 \vert \psi_1 \vert^2  \right ) \psi_2 \notag \\
&\hspace{1.2cm}+U_3 \psi_2^* \psi_1^2 - \frac{\Omega_0}{2} \psi_1 ,\label{NLS_effective_generalized}
\end{align}
where the three types of nonlinearities are controlled by the parameters $U_{1,2,3}$ displayed in Eq.~\eqref{g_def}. Consequently, the three types of effective nonlinearities can be tuned by adjusting the imbalanced pulse sequence. 

In an optics setting, the modification of (effective) optical nonlinearities will directly manifest in the topology of phase space [Fig.~\ref{fig_landscape_imbalanced}], which can be explored by extracting the trajectories $(z(t), \varphi(t))$ through light intensity and phase measurements.

\section{Concluding remarks}

This work proposed a method to engineer and tune nonlinearities in two-mode optical devices, using a designed pulse sequence that couples the optical modes in a fast and periodic manner. These repeated mixing operations simply correspond to the pulsed activation of a linear coupling between the two modes, and they can thus be implemented in a broad range of two-mode nonlinear systems, ranging from microresonators~\cite{cao2017experimental,cao2020reconfigurable} and two-waveguide couplers~\cite{szameit2010discrete,szameit2009inhibition} to circuit-QED platforms~\cite{roushan2017chiral}. While we considered a generic setting that includes both self-phase and cross-phase modulations in the absence of the periodic drive [Eq.~\eqref{NLS}], we found that effective nonlinearities emerge even when a single type of nonlinearity is present. Importantly, we demonstrated that the strength (and sign) of effective nonlinearities can be tuned by simply adjusting the pulse sequence. 

To detect the emergence of drive-induced nonlinearities, we proposed to study changes in the phase space's topology~\cite{zibold2010classical}, which can be explored by monitoring the dynamics of the relative intensity $z(t)$ and phase $\varphi(t)$ of the two optical modes. According to our numerical studies, these properties could already be revealed over ``time" scales of the order of $5-10T$, where $T$ denotes the period of the driving sequence. This is particularly appealing for waveguide settings~\cite{szameit2010discrete}, where the ``evolution time" associated with the propagation distance -- and hence the number of driving periods -- is limited. In this context, it would be interesting to combine such driving schemes with a state-recycling protocol~\cite{mukherjee2018state}.

While we considered a simple pulse sequence, characterized by the alternance of linear mixing operations and ``free" evolution, we note that more complicated protocols and configurations could be envisaged. For instance, different types of mixing processes could be activated within each period of the drive, including nonlinear processes. Moreover, we note that similar driving schemes could be designed for $N$-mode devices, such as realized in arrays of ultrafast-laser-inscribed waveguides~\cite{szameit2010discrete}. In the simplest scenario, one would couple the $N$ modes by pairs, i.e.~apply the pulse sequence in Eq.~\eqref{U_sequence} for the neighboring modes $1\leftrightarrow2, 3\leftrightarrow4, \dots , (N-1) \leftrightarrow N$, and then for the complementary pairs $2\leftrightarrow3, 4\leftrightarrow5, \dots , (N-2)\leftrightarrow (N-1)$, and repeat this whole sequence in a fast and time periodic manner. In this context, it would be exciting to study the interplay of drive-induced nonlinearities and topological band structures; see for instance Ref.~\cite{ivanov2021four}, where edge solitons were studied in the presence of four-wave mixing.

It would also be intriguing to explore the applicability of our scheme in the context of superconducting microwave cavities~\cite{chakram2021seamless}, where optical nonlinearities originate from the coupling to transmon ancillas. Indeed, it was recently shown that such optical nonlinearities can be modified by applying an off-resonant drive on the transmon ancillas~\cite{zhang2022drive}. Moreover, in circuit-QED platforms, the linear coupling between neighboring qubits can be modulated in a time-periodic manner~\cite{roushan2017chiral}; applying our pulse protocol to such settings could be used to modify the nonlinearity of the qubits, and hence, the interaction between microwave photons. In general, we anticipate that drive-induced nonlinearities, such as the effective four-wave mixing studied in this work, could be useful for nonlinear optics applications~\cite{agrawal2001applications,agrawal2012nonlinear}.

Finally, we remark that the present work relies on a non-dissipative theoretical framework, where the number of bosons is conserved. Our scheme could nevertheless be applied to driven-dissipative optical devices~\cite{carusotto2013quantum}, such as fiber ring cavities or microresonators described by the Lugiato-Lefever equation~\cite{lugiato1987spatial,haelterman1992dissipative,leo2010temporal,coen2013modeling,garbin2020asymmetric}, upon treating dissipation within the Floquet analysis ~\cite{higashikawa2018floquet,schnell2020there,schnell2021high}.

\paragraph*{Acknowledgments} 
This work was initiated through discussions with J. Fatome and S. Coen, who are warmly acknowledged. M.~Bukov, I.~Carusotto, N.~R. Cooper, A.~Eckardt, N.~Englebert, M.~Di Liberto, B.~Mera, F.~Petiziol and A.~Schnell are also acknowledged for various discussions. The author is also very grateful to S. Coen, N.~Englebert, J. Fatome, P.~Kockaert  and S. Mukherjee for their comments on the manuscript. The author is supported by the FRS-FNRS (Belgium), the ERC Starting Grant TopoCold and the EOS project CHEQS.

\begin{appendix}

\section{Useful formulas}\label{sect_app}

This work uses two families of operators:~the bosonic operators $\hat a_{1}^{(\dagger)}$ and $\hat a_{2}^{(\dagger)}$ associated with the two modes, and which satisfy the canonical bosonic commutation relations, $[\hat a_{\sigma},\hat a_{\sigma'}^{\dagger}]=\delta_{\sigma,\sigma'}$, where $\sigma\!=\!1,2$; and the angular momentum (Schwinger) operators, defined as
\begin{align}
&\hat J_x = \frac{1}{2} \left (\hat a_1^{\dagger} \hat a_2 + \hat a_2^{\dagger} \hat a_1 \right ) , \quad \hat J_y = \frac{1}{2i} \left (\hat a_2^{\dagger} \hat a_1 - \hat a_1^{\dagger} \hat a_2 \right ), \notag \\
&\hat J_z = \frac{1}{2} \left (\hat a_2^{\dagger} \hat a_2 - \hat a_1^{\dagger} \hat a_1 \right ) , \quad \hat N = \hat a_1^{\dagger} \hat a_1 + \hat a_2^{\dagger} \hat a_2. \label{app_Schwinger}
\end{align}
These operators satisfy the spin commutation relations $[\hat J_{\mu}, \hat J_{\nu}]=i\varepsilon_{\mu \nu \lambda} \hat J_{\lambda}$, and the operator $\hat N$ counts the total number of bosons in the system (assumed to be constant). 

In view of expressing interaction processes with Schwinger operators, it is useful to note that
\begin{align}
&\hat J_z^2 = \frac{1}{4} \left ( \hat a_1^{\dagger} \hat a_1^{\dagger} \hat a_1  \hat a_1 + \hat a_2^{\dagger} \hat a_2^{\dagger} \hat a_2 \hat a_2  - 2 \hat a_1^{\dagger} \hat a_2^{\dagger} \hat a_1  \hat a_2 + \hat N \right ) ,\notag\\
&\hat J_y^2 = \frac{1}{4} \left (2 \hat a_1^{\dagger} \hat a_2^{\dagger} \hat a_1  \hat a_2 - \hat a_1^{\dagger} \hat a_1^{\dagger} \hat a_2  \hat a_2 - \hat a_2^{\dagger} \hat a_2^{\dagger} \hat a_1 \hat a_1   + \hat N \right ) ,\notag \\
&\hat N^2 = \left (\hat a_1^{\dagger} \hat a_1^{\dagger} \hat a_1  \hat a_1 + \hat a_2^{\dagger} \hat a_2^{\dagger} \hat a_2 \hat a_2  + 2 \hat a_1^{\dagger} \hat a_2^{\dagger} \hat a_1  \hat a_2 + \hat N \right ) .\label{useful}
\end{align}
Hence, both $\hat J_z^2$ and $\hat N^2$ contain intra-mode (Hubbard) and inter-mode (cross) interactions, while $\hat J_y^2$ contains a combination of inter-mode interactions and pair-hopping processes [Fig.~\ref{fig_process}]. We point out that $\hat J_z^2$ is related to $\hat J_y^2$ through a unitary transformation; see Eq.~\eqref{BCH}.

From Eq.~\eqref{useful}, we can express the intra-mode (Hubbard) interaction terms as
\begin{equation}
\frac{1}{2} \left ( \hat a_1^{\dagger} \hat a_1^{\dagger} \hat a_1  \hat a_1 + \hat a_2^{\dagger} \hat a_2^{\dagger} \hat a_2 \hat a_2 \right ) = \hat J_z^2 + \text{constant},
\end{equation}
where the irrelevant constant term reads $\hat N (\hat N -2)/4$. Similarly, the inter-mode (cross) interaction term reads
\begin{equation}
\hat a_1^{\dagger} \hat a_2^{\dagger} \hat a_1  \hat a_2 = - \hat J_z^2 + \text{constant},
\end{equation}
with the irrelevant constant term $\hat N^2/4$. These expressions were used to derive the Hamiltonian in Eq.~\eqref{Josephson} from Eq.~\eqref{eq_parent_static}.

Finally, it is useful to note that a combination of intra-mode (Hubbard) interactions and pair-hopping processes can be expressed as
\begin{align}
\hat J_z^2 + \hat J_y^2&= \frac{1}{4} \left ( \hat a_1^{\dagger} \hat a_1^{\dagger} \hat a_1  \hat a_1 + \hat a_2^{\dagger} \hat a_2^{\dagger} \hat a_2 \hat a_2 \right ) \notag \\
&- \frac{1}{4} \left (\hat a_1^{\dagger} \hat a_1^{\dagger} \hat a_2  \hat a_2 + \hat a_2^{\dagger} \hat a_2^{\dagger} \hat a_1 \hat a_1 \right ) + \text{constant}.
\end{align}
This expression was used to derive the effective Hamiltonian in Eq.~\eqref{eff_bosonic} from Eq.~\eqref{eff_final}.

\end{appendix}


\begin{thebibliography}{100}

\bibitem{eckardt2017colloquium}
A.~Eckardt, ``Colloquium: Atomic quantum gases in periodically driven optical
  lattices,'' {\em Reviews of Modern Physics}, vol.~89, no.~1, p.~011004, 2017.

\bibitem{oka2019floquet}
T.~Oka and S.~Kitamura, ``Floquet engineering of quantum materials,'' {\em
  Annual Review of Condensed Matter Physics}, vol.~10, pp.~387--408, 2019.

\bibitem{rudner2020band}
M.~S. Rudner and N.~H. Lindner, ``Band structure engineering and
  non-equilibrium dynamics in floquet topological insulators,'' {\em Nature
  reviews physics}, vol.~2, no.~5, pp.~229--244, 2020.

\bibitem{weitenberg2021tailoring}
C.~Weitenberg and J.~Simonet, ``Tailoring quantum gases by floquet
  engineering,'' {\em Nature Physics}, vol.~17, no.~12, pp.~1342--1348, 2021.

\bibitem{georgescu2014quantum}
I.~M. Georgescu, S.~Ashhab, and F.~Nori, ``Quantum simulation,'' {\em Reviews
  of Modern Physics}, vol.~86, no.~1, p.~153, 2014.

\bibitem{altman2021quantum}
E.~Altman, K.~R. Brown, G.~Carleo, L.~D. Carr, E.~Demler, C.~Chin, B.~DeMarco,
  S.~E. Economou, M.~A. Eriksson, K.-M.~C. Fu, {\em et~al.}, ``Quantum
  simulators: Architectures and opportunities,'' {\em PRX Quantum}, vol.~2,
  no.~1, p.~017003, 2021.

\bibitem{salerno2016floquet}
G.~Salerno, T.~Ozawa, H.~M. Price, and I.~Carusotto, ``Floquet topological
  system based on frequency-modulated classical coupled harmonic oscillators,''
  {\em Physical Review B}, vol.~93, no.~8, p.~085105, 2016.

\bibitem{fleury2016floquet}
R.~Fleury, A.~B. Khanikaev, and A.~Alu, ``Floquet topological insulators for
  sound,'' {\em Nature communications}, vol.~7, no.~1, pp.~1--11, 2016.

\bibitem{rechtsman2013photonic}
M.~C. Rechtsman, J.~M. Zeuner, Y.~Plotnik, Y.~Lumer, D.~Podolsky, F.~Dreisow,
  S.~Nolte, M.~Segev, and A.~Szameit, ``Photonic floquet topological
  insulators,'' {\em Nature}, vol.~496, no.~7444, pp.~196--200, 2013.

\bibitem{schine2016synthetic}
N.~Schine, A.~Ryou, A.~Gromov, A.~Sommer, and J.~Simon, ``Synthetic landau
  levels for photons,'' {\em Nature}, vol.~534, no.~7609, pp.~671--675, 2016.

\bibitem{roushan2017chiral}
P.~Roushan, C.~Neill, A.~Megrant, Y.~Chen, R.~Babbush, R.~Barends, B.~Campbell,
  Z.~Chen, B.~Chiaro, A.~Dunsworth, {\em et~al.}, ``Chiral ground-state
  currents of interacting photons in a synthetic magnetic field,'' {\em Nature
  Physics}, vol.~13, no.~2, pp.~146--151, 2017.

\bibitem{ozawa2019topological}
T.~Ozawa, H.~M. Price, A.~Amo, N.~Goldman, M.~Hafezi, L.~Lu, M.~C. Rechtsman,
  D.~Schuster, J.~Simon, O.~Zilberberg, {\em et~al.}, ``Topological
  photonics,'' {\em Reviews of Modern Physics}, vol.~91, no.~1, p.~015006,
  2019.

\bibitem{goldman2014periodically}
N.~Goldman and J.~Dalibard, ``Periodically driven quantum systems: effective
  hamiltonians and engineered gauge fields,'' {\em Physical review X}, vol.~4,
  no.~3, p.~031027, 2014.

\bibitem{aidelsburger2018artificial}
M.~Aidelsburger, S.~Nascimbene, and N.~Goldman, ``Artificial gauge fields in
  materials and engineered systems,'' {\em Comptes Rendus Physique}, vol.~19,
  no.~6, pp.~394--432, 2018.

\bibitem{rapp2012ultracold}
{\'A}.~Rapp, X.~Deng, and L.~Santos, ``Ultracold lattice gases with
  periodically modulated interactions,'' {\em Physical review letters},
  vol.~109, no.~20, p.~203005, 2012.

\bibitem{ajoy2013quantum}
A.~Ajoy and P.~Cappellaro, ``Quantum simulation via filtered hamiltonian
  engineering: Application to perfect quantum transport in spin networks,''
  {\em Physical review letters}, vol.~110, no.~22, p.~220503, 2013.

\bibitem{di2014quantum}
M.~Di~Liberto, C.~E. Creffield, G.~Japaridze, and C.~M. Smith, ``Quantum
  simulation of correlated-hopping models with fermions in optical lattices,''
  {\em Physical Review A}, vol.~89, no.~1, p.~013624, 2014.

\bibitem{daley2014effective}
A.~J. Daley and J.~Simon, ``Effective three-body interactions via
  photon-assisted tunneling in an optical lattice,'' {\em Physical Review A},
  vol.~89, no.~5, p.~053619, 2014.

\bibitem{hung2016quantum}
C.-L. Hung, A.~Gonz{\'a}lez-Tudela, J.~I. Cirac, and H.~Kimble, ``Quantum spin
  dynamics with pairwise-tunable, long-range interactions,'' {\em Proceedings
  of the National Academy of Sciences}, vol.~113, no.~34, pp.~E4946--E4955,
  2016.

\bibitem{pieplow2018generation}
G.~Pieplow, F.~Sols, and C.~E. Creffield, ``Generation of atypical hopping and
  interactions by kinetic driving,'' {\em New Journal of Physics}, vol.~20,
  no.~7, p.~073045, 2018.

\bibitem{lee2018floquet}
C.~H. Lee, W.~W. Ho, B.~Yang, J.~Gong, and Z.~Papi{\'c}, ``Floquet mechanism
  for non-abelian fractional quantum hall states,'' {\em Physical review
  letters}, vol.~121, no.~23, p.~237401, 2018.

\bibitem{choi2020robust}
J.~Choi, H.~Zhou, H.~S. Knowles, R.~Landig, S.~Choi, and M.~D. Lukin, ``Robust
  dynamic hamiltonian engineering of many-body spin systems,'' {\em Physical
  Review X}, vol.~10, no.~3, p.~031002, 2020.

\bibitem{barbiero2020bose}
L.~Barbiero, L.~Chomaz, S.~Nascimbene, and N.~Goldman, ``Bose-hubbard physics
  in synthetic dimensions from interaction trotterization,'' {\em Physical
  Review Research}, vol.~2, no.~4, p.~043340, 2020.

\bibitem{dehghani2021light}
H.~Dehghani, M.~Hafezi, and P.~Ghaemi, ``Light-induced topological
  superconductivity via floquet interaction engineering,'' {\em Physical Review
  Research}, vol.~3, no.~2, p.~023039, 2021.

\bibitem{geier2021floquet}
S.~Geier, N.~Thaicharoen, C.~Hainaut, T.~Franz, A.~Salzinger, A.~Tebben,
  D.~Grimshandl, G.~Z{\"u}rn, and M.~Weidem{\"u}ller, ``Floquet hamiltonian
  engineering of an isolated many-body spin system,'' {\em Science}, vol.~374,
  no.~6571, pp.~1149--1152, 2021.

\bibitem{fausti2011light}
D.~Fausti, R.~Tobey, N.~Dean, S.~Kaiser, A.~Dienst, M.~C. Hoffmann, S.~Pyon,
  T.~Takayama, H.~Takagi, and A.~Cavalleri, ``Light-induced superconductivity
  in a stripe-ordered cuprate,'' {\em science}, vol.~331, no.~6014,
  pp.~189--191, 2011.

\bibitem{mitrano2016possible}
M.~Mitrano, A.~Cantaluppi, D.~Nicoletti, S.~Kaiser, A.~Perucchi, S.~Lupi,
  P.~Di~Pietro, D.~Pontiroli, M.~Ricc{\`o}, S.~R. Clark, {\em et~al.},
  ``Possible light-induced superconductivity in k3c60 at high temperature,''
  {\em Nature}, vol.~530, no.~7591, pp.~461--464, 2016.

\bibitem{struck2011quantum}
J.~Struck, C.~{\"O}lschl{\"a}ger, R.~Le~Targat, P.~Soltan-Panahi, A.~Eckardt,
  M.~Lewenstein, P.~Windpassinger, and K.~Sengstock, ``Quantum simulation of
  frustrated classical magnetism in triangular optical lattices,'' {\em
  Science}, vol.~333, no.~6045, pp.~996--999, 2011.

\bibitem{struck2013engineering}
J.~Struck, M.~Weinberg, C.~{\"O}lschl{\"a}ger, P.~Windpassinger, J.~Simonet,
  K.~Sengstock, R.~H{\"o}ppner, P.~Hauke, A.~Eckardt, M.~Lewenstein, {\em
  et~al.}, ``Engineering ising-xy spin-models in a triangular lattice using
  tunable artificial gauge fields,'' {\em Nature Physics}, vol.~9, no.~11,
  pp.~738--743, 2013.

\bibitem{gorg2018enhancement}
F.~G{\"o}rg, M.~Messer, K.~Sandholzer, G.~Jotzu, R.~Desbuquois, and
  T.~Esslinger, ``Enhancement and sign change of magnetic correlations in a
  driven quantum many-body system,'' {\em Nature}, vol.~553, no.~7689,
  pp.~481--485, 2018.

\bibitem{ponte2015many}
P.~Ponte, Z.~Papi{\'c}, F.~Huveneers, and D.~A. Abanin, ``Many-body
  localization in periodically driven systems,'' {\em Physical review letters},
  vol.~114, no.~14, p.~140401, 2015.

\bibitem{abanin2019colloquium}
D.~A. Abanin, E.~Altman, I.~Bloch, and M.~Serbyn, ``Colloquium: Many-body
  localization, thermalization, and entanglement,'' {\em Reviews of Modern
  Physics}, vol.~91, no.~2, p.~021001, 2019.

\bibitem{hensinger2001dynamical}
W.~K. Hensinger, H.~H{\"a}ffner, A.~Browaeys, N.~R. Heckenberg, K.~Helmerson,
  C.~McKenzie, G.~J. Milburn, W.~D. Phillips, S.~L. Rolston,
  H.~Rubinsztein-Dunlop, {\em et~al.}, ``Dynamical tunnelling of ultracold
  atoms,'' {\em Nature}, vol.~412, no.~6842, pp.~52--55, 2001.

\bibitem{arnal2020chaos}
M.~Arnal, G.~Chatelain, M.~Martinez, N.~Dupont, O.~Giraud, D.~Ullmo,
  B.~Georgeot, G.~Lemari{\'e}, J.~Billy, and D.~Gu{\'e}ry-Odelin,
  ``Chaos-assisted tunneling resonances in a synthetic floquet superlattice,''
  {\em Science advances}, vol.~6, no.~38, p.~eabc4886, 2020.

\bibitem{barbiero2019coupling}
L.~Barbiero, C.~Schweizer, M.~Aidelsburger, E.~Demler, N.~Goldman, and
  F.~Grusdt, ``Coupling ultracold matter to dynamical gauge fields in optical
  lattices: From flux attachment to ?2 lattice gauge theories,'' {\em Science
  advances}, vol.~5, no.~10, p.~eaav7444, 2019.

\bibitem{schweizer2019floquet}
C.~Schweizer, F.~Grusdt, M.~Berngruber, L.~Barbiero, E.~Demler, N.~Goldman,
  I.~Bloch, and M.~Aidelsburger, ``Floquet approach to $z_2$ lattice gauge
  theories with ultracold atoms in optical lattices,'' {\em Nature Physics},
  vol.~15, no.~11, pp.~1168--1173, 2019.

\bibitem{szameit2010discrete}
A.~Szameit and S.~Nolte, ``Discrete optics in femtosecond-laser-written
  photonic structures,'' {\em Journal of Physics B: Atomic, Molecular and
  Optical Physics}, vol.~43, no.~16, p.~163001, 2010.

\bibitem{mukherjee2017experimental}
S.~Mukherjee, A.~Spracklen, M.~Valiente, E.~Andersson, P.~{\"O}hberg,
  N.~Goldman, and R.~R. Thomson, ``Experimental observation of anomalous
  topological edge modes in a slowly driven photonic lattice,'' {\em Nature
  communications}, vol.~8, no.~1, pp.~1--7, 2017.

\bibitem{maczewsky2017observation}
L.~J. Maczewsky, J.~M. Zeuner, S.~Nolte, and A.~Szameit, ``Observation of
  photonic anomalous floquet topological insulators,'' {\em Nature
  communications}, vol.~8, no.~1, pp.~1--7, 2017.

\bibitem{mukherjee2018state}
S.~Mukherjee, H.~K. Chandrasekharan, P.~{\"O}hberg, N.~Goldman, and R.~R.
  Thomson, ``State-recycling and time-resolved imaging in topological photonic
  lattices,'' {\em Nature communications}, vol.~9, no.~1, pp.~1--6, 2018.

\bibitem{mukherjee2020observation}
S.~Mukherjee and M.~C. Rechtsman, ``Observation of floquet solitons in a
  topological bandgap,'' {\em Science}, vol.~368, no.~6493, pp.~856--859, 2020.

\bibitem{mukherjee2021observation}
S.~Mukherjee and M.~C. Rechtsman, ``Observation of unidirectional solitonlike
  edge states in nonlinear floquet topological insulators,'' {\em Physical
  Review X}, vol.~11, no.~4, p.~041057, 2021.

\bibitem{lustig2019photonic}
E.~Lustig, S.~Weimann, Y.~Plotnik, Y.~Lumer, M.~A. Bandres, A.~Szameit, and
  M.~Segev, ``Photonic topological insulator in synthetic dimensions,'' {\em
  Nature}, vol.~567, no.~7748, pp.~356--360, 2019.

\bibitem{mukherjee2018experimental}
S.~Mukherjee, M.~Di~Liberto, P.~{\"O}hberg, R.~R. Thomson, and N.~Goldman,
  ``Experimental observation of aharonov-bohm cages in photonic lattices,''
  {\em Physical review letters}, vol.~121, no.~7, p.~075502, 2018.

\bibitem{longhi2006observation}
S.~Longhi, M.~Marangoni, M.~Lobino, R.~Ramponi, P.~Laporta, E.~Cianci, and
  V.~Foglietti, ``Observation of dynamic localization in periodically curved
  waveguide arrays,'' {\em Physical review letters}, vol.~96, no.~24,
  p.~243901, 2006.

\bibitem{szameit2009inhibition}
A.~Szameit, Y.~V. Kartashov, F.~Dreisow, M.~Heinrich, T.~Pertsch, S.~Nolte,
  A.~T{\"u}nnermann, V.~A. Vysloukh, F.~Lederer, and L.~Torner, ``Inhibition of
  light tunneling in waveguide arrays,'' {\em Physical review letters},
  vol.~102, no.~15, p.~153901, 2009.

\bibitem{della2007visualization}
G.~Della~Valle, M.~Ornigotti, E.~Cianci, V.~Foglietti, P.~Laporta, and
  S.~Longhi, ``Visualization of coherent destruction of tunneling in an optical
  double well system,'' {\em Physical review letters}, vol.~98, no.~26,
  p.~263601, 2007.

\bibitem{mukherjee2015modulation}
S.~Mukherjee, A.~Spracklen, D.~Choudhury, N.~Goldman, P.~{\"O}hberg,
  E.~Andersson, and R.~R. Thomson, ``Modulation-assisted tunneling in
  laser-fabricated photonic wannier--stark ladders,'' {\em New journal of
  physics}, vol.~17, no.~11, p.~115002, 2015.

\bibitem{stutzer2018photonic}
S.~St{\"u}tzer, Y.~Plotnik, Y.~Lumer, P.~Titum, N.~H. Lindner, M.~Segev, M.~C.
  Rechtsman, and A.~Szameit, ``Photonic topological anderson insulators,'' {\em
  Nature}, vol.~560, no.~7719, pp.~461--465, 2018.

\bibitem{yuan2018synthetic}
L.~Yuan, Q.~Lin, M.~Xiao, and S.~Fan, ``Synthetic dimension in photonics,''
  {\em Optica}, vol.~5, no.~11, pp.~1396--1405, 2018.

\bibitem{dutt2019experimental}
A.~Dutt, M.~Minkov, Q.~Lin, L.~Yuan, D.~A. Miller, and S.~Fan, ``Experimental
  band structure spectroscopy along a synthetic dimension,'' {\em Nature
  communications}, vol.~10, no.~1, pp.~1--8, 2019.

\bibitem{dutt2020single}
A.~Dutt, Q.~Lin, L.~Yuan, M.~Minkov, M.~Xiao, and S.~Fan, ``A single photonic
  cavity with two independent physical synthetic dimensions,'' {\em Science},
  vol.~367, no.~6473, pp.~59--64, 2020.

\bibitem{balvcytis2021synthetic}
A.~Bal{\v{c}}ytis, T.~Ozawa, Y.~Ota, S.~Iwamoto, J.~Maeda, and T.~Baba,
  ``Synthetic dimension band structures on a si cmos photonic platform,'' {\em
  arXiv preprint arXiv:2105.13742}, 2021.

\bibitem{englebert2021bloch}
N.~Englebert, N.~Goldman, M.~Erkintalo, N.~Mostaan, S.-P. Gorza, F.~Leo, and
  J.~Fatome, ``Bloch oscillations of driven dissipative solitons in a synthetic
  dimension,'' {\em arXiv preprint arXiv:2112.10756}, 2021.

\bibitem{clark2019interacting}
L.~W. Clark, N.~Jia, N.~Schine, C.~Baum, A.~Georgakopoulos, and J.~Simon,
  ``Interacting floquet polaritons,'' {\em Nature}, vol.~571, no.~7766,
  pp.~532--536, 2019.

\bibitem{johansen2020multimode}
C.~H. Johansen, J.~Lang, A.~Morales, A.~Baumg{\"a}renter, T.~Donner, and
  F.~Piazza, ``Multimode-polariton superradiance via floquet engineering,''
  {\em arXiv preprint arXiv:2011.12309}, 2020.

\bibitem{shan2021giant}
J.-Y. Shan, M.~Ye, H.~Chu, S.~Lee, J.-G. Park, L.~Balents, and D.~Hsieh,
  ``Giant modulation of optical nonlinearity by floquet engineering,'' {\em
  Nature}, vol.~600, no.~7888, pp.~235--239, 2021.

\bibitem{zhang2022drive}
Y.~Zhang, J.~C. Curtis, C.~S. Wang, R.~Schoelkopf, and S.~Girvin,
  ``Drive-induced nonlinearities of cavity modes coupled to a transmon
  ancilla,'' {\em Physical Review A}, vol.~105, no.~2, p.~022423, 2022.

\bibitem{cao2017experimental}
Q.-T. Cao, H.~Wang, C.-H. Dong, H.~Jing, R.-S. Liu, X.~Chen, L.~Ge, Q.~Gong,
  and Y.-F. Xiao, ``Experimental demonstration of spontaneous chirality in a
  nonlinear microresonator,'' {\em Physical Review Letters}, vol.~118, no.~3,
  p.~033901, 2017.

\bibitem{hill2020effects}
L.~Hill, G.-L. Oppo, M.~T. Woodley, and P.~Del'Haye, ``Effects of self-and
  cross-phase modulation on the spontaneous symmetry breaking of light in ring
  resonators,'' {\em Physical Review A}, vol.~101, no.~1, p.~013823, 2020.

\bibitem{garbin2020asymmetric}
B.~Garbin, J.~Fatome, G.-L. Oppo, M.~Erkintalo, S.~G. Murdoch, and S.~Coen,
  ``Asymmetric balance in symmetry breaking,'' {\em Physical Review Research},
  vol.~2, no.~2, p.~023244, 2020.

\bibitem{andrews1997observation}
M.~Andrews, C.~Townsend, H.-J. Miesner, D.~Durfee, D.~Kurn, and W.~Ketterle,
  ``Observation of interference between two bose condensates,'' {\em Science},
  vol.~275, no.~5300, pp.~637--641, 1997.

\bibitem{smerzi1997quantum}
A.~Smerzi, S.~Fantoni, S.~Giovanazzi, and S.~Shenoy, ``Quantum coherent atomic
  tunneling between two trapped bose-einstein condensates,'' {\em Physical
  Review Letters}, vol.~79, no.~25, p.~4950, 1997.

\bibitem{zibold2010classical}
T.~Zibold, E.~Nicklas, C.~Gross, and M.~K. Oberthaler, ``Classical bifurcation
  at the transition from rabi to josephson dynamics,'' {\em Physical review
  letters}, vol.~105, no.~20, p.~204101, 2010.

\bibitem{pitaevskii2016bose}
L.~Pitaevskii and S.~Stringari, {\em Bose-Einstein condensation and
  superfluidity}, vol.~164.
\newblock Oxford University Press, 2016.

\bibitem{holthaus2001towards}
M.~Holthaus, ``Towards coherent control of a bose-einstein condensate in a
  double well,'' {\em Physical Review A}, vol.~64, no.~1, p.~011601, 2001.

\bibitem{kramer2005parametric}
M.~Kr{\"a}mer, C.~Tozzo, and F.~Dalfovo, ``Parametric excitation of a
  bose-einstein condensate in a one-dimensional optical lattice,'' {\em
  Physical Review A}, vol.~71, no.~6, p.~061602, 2005.

\bibitem{susanto2008effects}
H.~Susanto, P.~Kevrekidis, B.~Malomed, and F.~K. Abdullaev, ``Effects of
  time-periodic linear coupling on two-component bose--einstein condensates in
  two dimensions,'' {\em Physics Letters A}, vol.~372, no.~10, pp.~1631--1638,
  2008.

\bibitem{lellouch2017parametric}
S.~Lellouch, M.~Bukov, E.~Demler, and N.~Goldman, ``Parametric instability
  rates in periodically driven band systems,'' {\em Physical Review X}, vol.~7,
  no.~2, p.~021015, 2017.

\bibitem{driben2018nonlinearity}
R.~Driben, V.~Konotop, B.~Malomed, T.~Meier, and A.~Yulin,
  ``Nonlinearity-induced localization in a periodically driven semidiscrete
  system,'' {\em Physical Review E}, vol.~97, no.~6, p.~062210, 2018.

\bibitem{kidd2019quantum}
R.~Kidd, M.~Olsen, and J.~Corney, ``Quantum chaos in a bose-hubbard dimer with
  modulated tunneling,'' {\em Physical Review A}, vol.~100, no.~1, p.~013625,
  2019.

\bibitem{maczewsky2020nonlinearity}
L.~J. Maczewsky, M.~Heinrich, M.~Kremer, S.~K. Ivanov, M.~Ehrhardt,
  F.~Martinez, Y.~V. Kartashov, V.~V. Konotop, L.~Torner, D.~Bauer, {\em
  et~al.}, ``Nonlinearity-induced photonic topological insulator,'' {\em
  Science}, vol.~370, no.~6517, pp.~701--704, 2020.

\bibitem{mochizuki2021fate}
K.~Mochizuki, K.~Mizuta, and N.~Kawakami, ``Fate of topological edge states in
  disordered periodically driven nonlinear systems,'' {\em Physical Review
  Research}, vol.~3, no.~4, p.~043112, 2021.

\bibitem{ivanov2021topological}
S.~K. Ivanov, Y.~V. Kartashov, M.~Heinrich, A.~Szameit, L.~Torner, and V.~V.
  Konotop, ``Topological dipole floquet solitons,'' {\em Physical Review A},
  vol.~103, no.~5, p.~053507, 2021.

\bibitem{higashikawa2018floquet}
S.~Higashikawa, H.~Fujita, and M.~Sato, ``Floquet engineering of classical
  systems,'' {\em arXiv preprint arXiv:1810.01103}, 2018.

\bibitem{sentef2020quantum}
M.~A. Sentef, J.~Li, F.~K{\"u}nzel, and M.~Eckstein, ``Quantum to classical
  crossover of floquet engineering in correlated quantum systems,'' {\em
  Physical Review Research}, vol.~2, no.~3, p.~033033, 2020.

\bibitem{carusotto2013quantum}
I.~Carusotto and C.~Ciuti, ``Quantum fluids of light,'' {\em Reviews of Modern
  Physics}, vol.~85, no.~1, p.~299, 2013.

\bibitem{cao2020reconfigurable}
Q.-T. Cao, R.~Liu, H.~Wang, Y.-K. Lu, C.-W. Qiu, S.~Rotter, Q.~Gong, and Y.-F.
  Xiao, ``Reconfigurable symmetry-broken laser in a symmetric microcavity,''
  {\em Nature communications}, vol.~11, no.~1, pp.~1--7, 2020.

\bibitem{goldman2015periodically}
N.~Goldman, J.~Dalibard, M.~Aidelsburger, and N.~R. Cooper, ``Periodically
  driven quantum matter: The case of resonant modulations,'' {\em Physical
  Review A}, vol.~91, no.~3, p.~033632, 2015.

\bibitem{bukov2015universal}
M.~Bukov, L.~D'Alessio, and A.~Polkovnikov, ``Universal high-frequency behavior
  of periodically driven systems: from dynamical stabilization to floquet
  engineering,'' {\em Advances in Physics}, vol.~64, no.~2, pp.~139--226, 2015.

\bibitem{eckardt2015high}
A.~Eckardt and E.~Anisimovas, ``High-frequency approximation for periodically
  driven quantum systems from a floquet-space perspective,'' {\em New journal
  of physics}, vol.~17, no.~9, p.~093039, 2015.

\bibitem{mikami2016brillouin}
T.~Mikami, S.~Kitamura, K.~Yasuda, N.~Tsuji, T.~Oka, and H.~Aoki,
  ``Brillouin-wigner theory for high-frequency expansion in periodically driven
  systems: Application to floquet topological insulators,'' {\em Physical
  Review B}, vol.~93, no.~14, p.~144307, 2016.

\bibitem{kockaert2006fast}
P.~Kockaert, C.~Cambournac, M.~Haelterman, G.~Kozyreff, and T.~Erneux, ``Fast
  self-pulsing through nonlinear incoherent feedback,'' {\em Optics letters},
  vol.~31, no.~4, pp.~495--497, 2006.

\bibitem{kozyreff2006fast}
G.~Kozyreff, T.~Erneux, M.~Haelterman, and P.~Kockaert, ``Fast optical
  self-pulsing in a temporal analog of the kerr-slice pattern-forming system,''
  {\em Physical Review A}, vol.~73, no.~6, p.~063815, 2006.

\bibitem{fatome2021self}
J.~Fatome, G.~Xu, B.~Garbin, N.~Berti, G.-L. Oppo, S.~G. Murdoch, M.~Erkintalo,
  and S.~Coen, ``Self-symmetrization of symmetry-breaking dynamics in passive
  kerr resonators,'' {\em arXiv preprint arXiv:2106.07642}, 2021.

\bibitem{gardiner2014quantum}
C.~Gardiner and P.~Zoller, {\em The Quantum World of Ultra-Cold Atoms and Light
  Book I: Foundations of Quantum Optics}, vol.~2.
\newblock World Scientific Publishing Company, 2014.

\bibitem{kitagawa1993squeezed}
M.~Kitagawa and M.~Ueda, ``Squeezed spin states,'' {\em Physical Review A},
  vol.~47, no.~6, p.~5138, 1993.

\bibitem{gross2010nonlinear}
C.~Gross, T.~Zibold, E.~Nicklas, J.~Esteve, and M.~K. Oberthaler, ``Nonlinear
  atom interferometer surpasses classical precision limit,'' {\em Nature},
  vol.~464, no.~7292, pp.~1165--1169, 2010.

\bibitem{duncan2020synthetic}
C.~W. Duncan, M.~J. Hartmann, R.~R. Thomson, and P.~{\"O}hberg, ``Synthetic
  mean-field interactions in photonic lattices,'' {\em The European Physical
  Journal D}, vol.~74, no.~5, pp.~1--7, 2020.

\bibitem{kraych2019nonlinear}
A.~E. Kraych, P.~Suret, G.~El, and S.~Randoux, ``Nonlinear evolution of the
  locally induced modulational instability in fiber optics,'' {\em Physical
  review letters}, vol.~122, no.~5, p.~054101, 2019.

\bibitem{kraych2019statistical}
A.~E. Kraych, D.~Agafontsev, S.~Randoux, and P.~Suret, ``Statistical properties
  of the nonlinear stage of modulation instability in fiber optics,'' {\em
  Physical review letters}, vol.~123, no.~9, p.~093902, 2019.

\bibitem{kraych2020instabilites}
A.~Kraych, {\em Instabilit{\'e}s modulationnelles dans un anneau de
  recirculation fibr{\'e}}.
\newblock PhD thesis, Lille, 2020.

\bibitem{schumm2005bose}
T.~Schumm, {\em Bose-Einstein condensates in magnetic double well potentials}.
\newblock PhD thesis, 2005.

\bibitem{agrawal2001applications}
G.~Agrawal, {\em Applications of nonlinear fiber optics}.
\newblock Elsevier, 2001.

\bibitem{agrawal2012nonlinear}
G.~Agrawal, {\em Nonlinear fiber optics}.
\newblock Elsevier, 2012.

\bibitem{footnote}
 {\em The nonlinear Schr\"odinger equation in Eq.~\eqref{NLS} assumes that the
  parameter $\gamma$ is the same for both modes. If this is not the case, then
  Eq.~\eqref{NLS} would display a term of the form $\sim (- \partial^2/\partial
  x^2) \hat \sigma_z$. Since the Pauli matrix $\hat \sigma_z$ does not commute
  with the mixing operator $\hat \sigma_x$ in Eq.~\eqref{pi_over_two}, this
  kinetic term will be modified in the effective Hamiltonian
  [Eq.~\eqref{NLS_effective}].}

\bibitem{bloch2008many}
I.~Bloch, J.~Dalibard, and W.~Zwerger, ``Many-body physics with ultracold
  gases,'' {\em Reviews of modern physics}, vol.~80, no.~3, p.~885, 2008.

\bibitem{dutta2015non}
O.~Dutta, M.~Gajda, P.~Hauke, M.~Lewenstein, D.-S. L{\"u}hmann, B.~A. Malomed,
  T.~Sowi{\'n}ski, and J.~Zakrzewski, ``Non-standard hubbard models in optical
  lattices: a review,'' {\em Reports on Progress in Physics}, vol.~78, no.~6,
  p.~066001, 2015.

\bibitem{auerbach2012interacting}
A.~Auerbach, {\em Interacting electrons and quantum magnetism}.
\newblock Springer Science \& Business Media, 2012.

\bibitem{zibold2012classical}
T.~Zibold, {\em Classical bifurcation and entanglement generation in an
  internal bosonic josephson junction}.
\newblock PhD thesis, 2012.

\bibitem{lipkin1965validity}
H.~J. Lipkin, N.~Meshkov, and A.~Glick, ``Validity of many-body approximation
  methods for a solvable model:(i). exact solutions and perturbation theory,''
  {\em Nuclear Physics}, vol.~62, no.~2, pp.~188--198, 1965.

\bibitem{paraoanu2001josephson}
G.-S. Paraoanu, S.~Kohler, F.~Sols, and A.~Leggett, ``The josephson plasmon as
  a bogoliubov quasiparticle,'' {\em Journal of Physics B: Atomic, Molecular
  and Optical Physics}, vol.~34, no.~23, p.~4689, 2001.

\bibitem{di2019nonlinear}
M.~Di~Liberto, S.~Mukherjee, and N.~Goldman, ``Nonlinear dynamics of
  aharonov-bohm cages,'' {\em Physical Review A}, vol.~100, no.~4, p.~043829,
  2019.

\bibitem{kitagawa2010topological}
T.~Kitagawa, E.~Berg, M.~Rudner, and E.~Demler, ``Topological characterization
  of periodically driven quantum systems,'' {\em Physical Review B}, vol.~82,
  no.~23, p.~235114, 2010.

\bibitem{anisimovas2015role}
E.~Anisimovas, G.~{\v{Z}}labys, B.~M. Anderson, G.~Juzeli{\=u}nas, and
  A.~Eckardt, ``Role of real-space micromotion for bosonic and fermionic
  floquet fractional chern insulators,'' {\em Physical Review B}, vol.~91,
  no.~24, p.~245135, 2015.

\bibitem{gajda1997fluctuations}
M.~Gajda and K.~Rzazewski, ``Fluctuations of bose-einstein condensate,''
  {\em Physical review letters}, vol.~78, no.~14, p.~2686, 1997.

\bibitem{larre2015propagation}
P.-{\'E}. Larr{\'e} and I.~Carusotto, ``Propagation of a quantum fluid of light
  in a cavityless nonlinear optical medium: General theory and response to
  quantum quenches,'' {\em Physical Review A}, vol.~92, no.~4, p.~043802, 2015.

\bibitem{julia2012dynamic}
B.~Julia-Diaz, T.~Zibold, M.~Oberthaler, M.~Mele-Messeguer, J.~Martorell, and
  A.~Polls, ``Dynamic generation of spin-squeezed states in bosonic josephson
  junctions,'' {\em Physical Review A}, vol.~86, no.~2, p.~023615, 2012.

\bibitem{bruno2012quantum}
P.~Bruno, ``Quantum geometric phase in majorana’s stellar representation:
  mapping onto a many-body aharonov-bohm phase,'' {\em Physical Review
  Letters}, vol.~108, no.~24, p.~240402, 2012.

\bibitem{strobel2014fisher}
H.~Strobel, W.~Muessel, D.~Linnemann, T.~Zibold, D.~B. Hume, L.~Pezz{\`e},
  A.~Smerzi, and M.~K. Oberthaler, ``Fisher information and entanglement of
  non-gaussian spin states,'' {\em Science}, vol.~345, no.~6195, pp.~424--427,
  2014.

\bibitem{evrard2019enhanced}
A.~Evrard, V.~Makhalov, T.~Chalopin, L.~A. Sidorenkov, J.~Dalibard, R.~Lopes,
  and S.~Nascimbene, ``Enhanced magnetic sensitivity with non-gaussian quantum
  fluctuations,'' {\em Physical review letters}, vol.~122, no.~17, p.~173601,
  2019.

\bibitem{nascimbene2020quantum}
S.~Nascimbene, ``Quantum-enhanced sensing and topological matter with ultracold
  dysprosium atoms,'' 2020.

\bibitem{ivanov2021four}
S.~K. Ivanov, Y.~V. Kartashov, and V.~V. Konotop, ``Four-wave mixing floquet
  topological solitons,'' {\em Optics Letters}, vol.~46, no.~19,
  pp.~4710--4713, 2021.

\bibitem{chakram2021seamless}
S.~Chakram, A.~E. Oriani, R.~K. Naik, A.~V. Dixit, K.~He, A.~Agrawal, H.~Kwon,
  and D.~I. Schuster, ``Seamless high-q microwave cavities for multimode
  circuit quantum electrodynamics,'' {\em Physical review letters}, vol.~127,
  no.~10, p.~107701, 2021.

\bibitem{lugiato1987spatial}
L.~A. Lugiato and R.~Lefever, ``Spatial dissipative structures in passive
  optical systems,'' {\em Physical review letters}, vol.~58, no.~21, p.~2209,
  1987.

\bibitem{haelterman1992dissipative}
M.~Haelterman, S.~Trillo, and S.~Wabnitz, ``Dissipative modulation instability
  in a nonlinear dispersive ring cavity,'' {\em Optics communications},
  vol.~91, no.~5-6, pp.~401--407, 1992.

\bibitem{leo2010temporal}
F.~Leo, S.~Coen, P.~Kockaert, S.-P. Gorza, P.~Emplit, and M.~Haelterman,
  ``Temporal cavity solitons in one-dimensional kerr media as bits in an
  all-optical buffer,'' {\em Nature Photonics}, vol.~4, no.~7, pp.~471--476,
  2010.

\bibitem{coen2013modeling}
S.~Coen, H.~G. Randle, T.~Sylvestre, and M.~Erkintalo, ``Modeling of
  octave-spanning kerr frequency combs using a generalized mean-field
  lugiato--lefever model,'' {\em Optics letters}, vol.~38, no.~1, pp.~37--39,
  2013.

\bibitem{schnell2020there}
A.~Schnell, A.~Eckardt, and S.~Denisov, ``Is there a floquet lindbladian?,''
  {\em Physical Review B}, vol.~101, no.~10, p.~100301, 2020.

\bibitem{schnell2021high}
A.~Schnell, S.~Denisov, and A.~Eckardt, ``High-frequency expansions for
  time-periodic lindblad generators,'' {\em Physical Review B}, vol.~104,
  no.~16, p.~165414, 2021.

\end{thebibliography}


\end{document}